%% file: 1481.TEX
\documentclass[]{aa}
\usepackage{graphicx}
\usepackage{txfonts}
\begin{document}

\title{ Filtration of atmospheric noise in narrow-field astrometry
	with very large telescopes}

\author{P.F.Lazorenko 
        	\inst{1}
	\and G.A.Lazorenko
	\inst{1}}

   \offprints{P.F.Lazorenko}

     \institute{Main Astronomical Observatory,
          National Academy of Sciencies of the Ukraine,
             Zabolotnogo 27, 03680 Kyiv-127, Ukraine\\
             email: laz@mao.kiev.ua  }

   \date{Received 24 June 2004; accepted 19 July 2004}                                                            

   \abstract{ This paper presents a non-classic approach to 
narrow field astrometry that offers a significant improvement over
conventional techniques due to enhanced 
reduction of atmospheric image motion.
The method is based on two key elements: apodization of the
entrance pupil and the enhanced virtual symmetry of reference stars.
Symmetrization is implemented by setting special weights
to each reference star. Thus a reference field itself forms a virtual
net filter that effectively attenuates the image motion spectrum.
Atmospheric positional error  was found to follow
a power dependency $\Delta \sim R^{k \mu/2} D^{-k/2+1/3}$ on angular
field size $R$ and aperture $D$; here $k$ is some optional even
integer $2 \leq k \leq \sqrt{8N+1}-1$ limited by a number $N$ of
reference stars, and $\mu \leq 1$ is a term dependent on $k$ and
the magnitude and sky star distribution in the field.
As compared to conventional techniques for which $k=2$, the improvement
in accuracy increases by some orders.
Limitations to astrometric performance of monopupil
large ground-based telescopes are estimated.  The total 
atmospheric and photon noise for
at a 10 m telescope at good $0.4''$ seeing was found to be,
depending on sky star density,
10 to 60~$\mu$as per 10~min exposure in R band. For a 100~m telescope 
and FWHM=0.1$''$ (low-order adaptive optics corrections) the potential
accuracy is 0.2 to 2~$\mu$as.

      \keywords{ atmospheric effects -- turbulence --
              methods: data analysis --
              astrometry }
              }

  \titlerunning{Filtration of atmospheric noise} %in narrow-field astrometry}
  \authorrunning{Lazorenko P. \& Lazorenko G.}

     \maketitle

\input 1481_1
\input 1481_2
\input 1481_3

\input 1481_4
\input 1481_5

\input 1481_6
\input 1481_7

\input 1481_8
\input 1481_9

\input ref.tex
\end{document}

%% file: 1481_1.tex
\section{Introduction}
Atmospheric turbulence affects astronomical observations with respect to
both image quality degradation and random image motion.
With improvement of image detectors
the second source of errors has become dominant in astrometric measurements.
The very slow progress in overcoming  atmospheric errors finally turned to 
development of high-accurate 
astrometry from space missions. The first successful Hipparcos
mission yielded data with an accuracy of 1 mas, the GAIA project
(Perryman et all. 2001) is aimed at 10 $\mu$as accuracy level. 

The most promising direction of optical ground-based astrometry is now 
related to the infrared long-baseline interferometers; the atmospheric
limit for this kind of instruments is about 10 $\mu$as (Shao \& 
Colavita 1992; Paresce et all. 2002). Astrometry with monopupil long-focus
telescopes is limited by much larger $\Delta \sim 1$ mas atmospheric 
error,
which was confirmed by numerous experimental and theoretical data.
Thus, Gatewood (1987), on the basis of measurements with the Multichannel
Astrometric Photometer ($D=0.76$ m), derived  $\Delta \simeq 3$ mas/hr 
with a reference frame of $10-20'$. Using the same data, Han (1989)
suggested a model of a image motion power spectrum according to  which
the distance between a $1'$ separated double star can be measured with an
accuracy of $ \simeq 1$ mas/1hr.
The best precision of 150 $\mu$as/hr ever obtained 
with ground-based telescopes was demonstrated by Pravdo \& Shaklan 
(1996) with the 5-m Palomar telescope in a field of 90 arcsec. 

A theoretical aspect of the image motion problem was studied by Lindegren
(1980) who derived analytical expressions for relative image motion
of a double star. For very narrow fields, 
 $\Delta^2 \sim 
(\rho h)^2 D^{-4/3}T^{-1}$ demonstrated  a rather weak
power dependency  of  the variance of image motion $\Delta^2$ on
the double star angular separation $\rho $, 
 objective diameter $D$ and   exposure $T$ ($h$ is the turbulent layer 
height). 
The above dependence shows that an improvement of $\Delta$ from 1 mas to
say 10 $\mu$as will require unrealistically large 
$D$ and $T$, unless $\rho $ is  limited to uselessly small angles.
The situation is somewhat better when a very dense star field is used
as a reference. For circular star samples of a radius $\rho$
the dependency becomes $\Delta^2 \sim 
(\rho  h)^{8/3} D^{-2}T^{-1}$ which is a factor $(\rho h/ D)^{2/3}$
 better
than that valid for a 
single reference star. Lazorenko (2002a), making allowance 
for
the second order term of the random phase fluctuations over the pupil,
has proved that the  dependency is $\Delta^2 \sim 
(\rho  h)^{11/3} D^{-3}T^{-1}$,
or a factor $\rho h/D$ still stronger.  
He also suggested a symmetrizing procedure by means of which 
any arbitrary distributed sample of stars can be utilized as 
a virtually symmetric reference frame. 

Nevertheless even the use of symmetric (or symmetrized) reference fields 
not can provide accuracies like those expected from
interferometric and/or space mission techniques. 
Thus, Lazorenko (2002a), considering a 10 m telescope, $1'$ diameter 
reference frame and all the turbulence concentrated at 2.8 km height,
found that atmospheric noise is 5 $\mu$as/hr at moderate seeing.
Scaling this data  to more realistic 
effective heights of 14 -- 17 km yields $\Delta = 120$ -- 300 $\mu$as/hr,
which is the limit of this technique. 
An efficient solution of the problem suggested in the current paper is
based on enhanced symmetrization of reference fields, which
is an improved version of the former procedure  (Lazorenko 2002a).
This new approach  allows a
much better filtration of random wave-front distortions,
reducing atmospheric noise to 1--10 $\mu$as/hr with
a 10 m aperture.

This paper presents the analysis of atmospheric limits to astrometric
performance of one-aperture telescopes. 
We emphasise the use of very large 10 -- 100 m 
telescopes since we want to obtain  limiting estimates of 
ground-based astrometry. Astrophysical drives and possible 
applications for microarcsecond astrometry are not discussed here
although they were considered in some papers (Perryman et al. 2001;
Pravdo \& Shaklan 1996; Paresce et al. 2002). 
In Section 2  we describe a model of a
differential image motion spectrum. The image motion
spectrum filtration  described in Section 3 is based on
two key elements: apodization of the entrance pupil (Section 4)
and the enhanced virtual symmetry of reference groups (Section 5).
In Section 6 we consider asymptotic properties of high-order
symmetry dense reference fields  and
in  Section 7 we discuss modifications of plate reductions 
necessary to implement the new technique. Estimates of the image
motion error integrated over the atmosphere are given in Section 8. 
Astrometric performance of very large telescopes under assumption of a 
two-component error budget is analyzed in Section 9.

%% file: 1481_2.tex
\section {A model of differential image motion  power spectrum}
Atmospheric random motion of the star image centroid
is caused by turbulence which is known to be
concentrated in a limited number of thin atmospheric
layers.
Each layer is described by its height $h$ above the ground,
a thickness $\Delta h$ which is typically 10 -- 100 m, rarely up to 800 m 
(e.g. Barletti et al. 1977; Redfern 1991; Vinnichenko et al. 1976)
and a refractive-index structure constant $C^2_n$. 
Turbulent motion in each layer is isotropic in horizontal
directions, and  refractive-index fluctuations are 
described by the structure function
$D_n({\bf r})=C^2_n r^p$,
where $\bf r$ is a vector connecting two points in a 2-D layer's
plane, and $p$ is a constant.  Some experimental and theoretical studies 
discussed by Lazorenko (2002a) testify that 
$p$ is to be considered as a variable term with a typical range of 
variations 1/2 -- 2/3 which at some particular atmospheric conditions is
widened from -1/3 to +1. In 
the present study, for the sake of universality  of
the derived equations, this quantity is considered as a variable, but
numerical estimates are made with $p=2/3$ only. 

Due to the small thickness of layers $\Delta h \ll h$, the each one
is considered as a thin horizontal  phase screen 
that produces random phase distortions of the light 
wave front.  
A  plane  wave that crosses the screen thus gains
some  random phase $\phi (x,y)$ where $x, y$ are Cartesian
coordinates in the phase screen plane.
Given the assumption of homogeneity and isotropy of turbulence
(at least in a horizontal plane),
the power spectral  density of a phase is a function 
\begin{equation}
\label{eq:2_2}
\hat{F}_{\phi}(q)=c_{\phi}q^{-p-3}
\end{equation}
of  a circular  spatial frequency  $q$.
Here  $c_{\phi}$ is a factor accounting for  turbulence  strength. 
In case of the Kolmogorov turbulence, $c_{\phi}$ is related to
$C^2_n$ via the expression (Tatarsky 1961)
\begin{equation}
\label{eq:2_3}
c_{\phi} =0.033 (2\pi /\lambda)^2 (2\pi)^{-2/3}C^2_n\Delta h
\end{equation}
where $\lambda$ is wavelength.

Perturbations of a light phase  $\phi (x,y)$ lead to
random displacements $\zeta$ of star image centroids in  the focal 
plane. 
The  magnitude of $\zeta$ measured in some arbitrary
direction $\bf{r'}$  is proportional to the  gradient of the
phase $\partial \phi (x,y) / \partial \bf{r'}$
averaged over the screen section involved
in  image formation. For the instant exposure $T=0$ it is
a circular area of a diameter $D$ equal to the telescope diameter and
centered at the point $\bf r$  
where the phase screen is crossed by light
beams passing from  the star to the centre of
the telescope pupil. A quantity $\zeta$ as a 
function of $\bf r$ vector is given by the convolution 
(Martin 1987; Conan et al. 1995) 
\begin{equation}
\label{eq:2_4}
\zeta ({\bf r})=\frac{\lambda }{2\pi }
\frac{\partial \phi ({\bf r})}{\partial {\bf r' }}
\ast \frac{P({\bf r})}{\sigma} 
\end{equation}
where $P({\bf r})$ is the entrance pupil function that 
determines optical transmission of the objective and 
\begin{equation}
\label{eq:2_6}
\sigma  = \int \int P({\bf r}) \, d{\bf r} 
\end{equation}
is the effective area of the entrance pupil.
For a classical  fully transparent monopupil
\begin{equation}
\label{eq:2_5}
P({\bf r})=\left \{ \begin{array}{ll}
1, & r \leq D/2\\
0,         & r > D/2
\end{array} \right.
\end{equation}

At finite exposure $T>0$, the wind-induced motion 
of a phase screen with the wind velocity $V$ is to be taken into 
consideration since it causes extra  averaging of phase
fluctuations along a straight line of length $VT$. 
This effect is described by
convolution of the function (\ref{eq:2_4}) with a rectangular
function
\begin{equation}
\label{eq:2_7}
\mbox{rect}(z)=\left \{ \begin{array}{ll}
(VT)^{-1}, & z \leq VT/2\\
0,         & z > VT/2
\end{array} \right.
\end{equation}
where $z$ is measured along the  wind direction which
forms some angle $\varepsilon$ with the $x$-axis.

\begin{figure}[tbh]  %fig. 1
%\label{fig:geom}
\includegraphics*[ width=8.3cm, height=6.2cm]{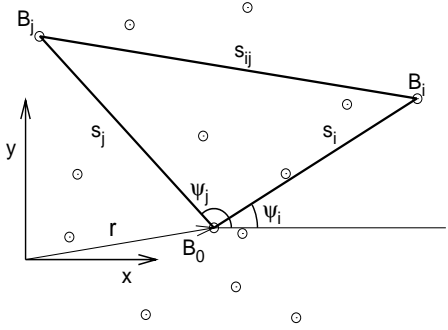}
\caption{Geometric elements specifying a star group (circles) projection on a 
phase screen}
\end{figure}

In differential astrometry the position of the
target object is measured with respect to some $N$ reference stars.
Let us consider a geometry of  a stellar group in  projection  
on the phase screen displayed in Fig.1. The points B$_i$, 
$i=1 \ldots N$ represent reference stars,  the
target star projection  B$_0$ is given by the vector ${\bf r}$.  
Cartesian coordinates $x_i$, $y_i$ of stars are  equal
to their standard coordinates scaled to the layer height $h$.
In polar coordinates, the distribution of points B$_i$ is given by their 
linear separations from the target $s_i=h \rho_i$ where
$\rho_i$ is $i$-th star
angular separation from the target, by positional angles 
$\psi_i$ measured with respect to the $x$-axis and
by distances
$s_{ij}=\sqrt{s_i^2+s_j^2-2s_i s_j \cos(\psi_i-\psi_j)}$
between the two points B$_i$ and B$_j$.

Differential displacement of star images  $\Delta \zeta$ is
obviously equal to the difference of $\zeta ({\bf r})$ function values
taken at points B$_0$ and B$_i$. In the trivial case of one reference star 
B$_1$, the differential displacement is 
$\Delta \zeta ({\bf r})=\zeta ({\bf r})-\zeta ({\bf r+s}_1)=
\zeta ({\bf r}) \ast \tilde{Q}({\bf r})$
where
\begin{equation}
\label{eq:2_8a}
\tilde{Q}({\bf r})=\delta ({\bf r})-\delta ({\bf r+s}_1)
\end{equation}
is a difference operator. For multi-star reference frames the form of the
function (\ref{eq:2_8a})  is  specified by distribution of 
reference stars which we include by using designation
$\tilde{Q}({\bf r,s})$.

Taking into account the above effects, we come to a common expression
\begin{equation}
\label{eq:2_8}
\Delta \zeta ({\bf r})=\frac{\lambda }{2\pi }
\frac{\partial \phi ({\bf r})}{\partial {\bf r'} }\ast
\mbox{rect} ({\bf r}) 
\ast P({\bf r})/ \sigma \ast \tilde{Q}({\bf r,s})
\end{equation}
inherent to the differential technique. 

Now it is useful  to proceed to the analysis in the frequency domain
where the power spectral density $F_{\zeta}(q,\varphi)$ 
of the quantity $\Delta \zeta$ is  expressed as  
a product of  four squared  Fourier transforms ${\mathcal F}{ }$ 
of  functions convolved in (\ref{eq:2_8}) and  the power spectrum 
(\ref{eq:2_2}). In the polar coordinates $q$, $\varphi$
\begin{equation}
\label{eq:2_9}
\begin{array}{ll}
F_{\zeta}(q,\varphi )=&   [\lambda/(2\pi)]^2  
 {\mathcal F}^2 \left \{ \frac{\partial}{\partial {\bf r'}} \right \} 
{\mathcal F}^2 \left \{ \frac{P}{\sigma} \right \} 
{\mathcal F}^2 \{\mbox{rect} \} \\
	& \times {\mathcal F}^2 \{ \tilde{Q} \} F_{\phi}(q)\\
\end{array}
\end{equation}

It is known that $ {\mathcal F}^2 
 \{ \partial / \partial {\bf r'}\} =[2\pi q \mbox{cos}
(\varphi-\theta)]^2$ and  $ {\mathcal F}^2 \{ \mbox{rect} 
\}=\mbox{sinc}^2
[\pi V T q \mbox{cos}(\varphi-\varepsilon)]$ where $ \theta $ is the 
angle formed by the vector ${\bf r'}$ and $x$-axis and 
sinc$ (z) =\sin (z) / z$ (e.g. Martin 1987; Lazorenko 2002a). 
Angular parameters  $\varepsilon$ (the wind direction) and $\theta$ 
 (direction of image motion measurements) present in 
the above
expressions introduce anisotropic effects (Lazorenko 2002a), 
which makes the analysis rather complicated and requires a special 
discussion. 
However, there are some reasons permitting us to  perform averaging
of Eq.(\ref{eq:2_9}) over  $\varepsilon$ and $\theta$.
With respect to $\theta$, this procedure is justified by the
fact that, normally, the image motion variance is defined as its 
mean value measured along the $x$ and $y$ axes, hence 
${\mathcal F}^2 \{ \partial/ \partial {\bf r'}\} =2\pi^2 q^2$. 
Averaging over another angle $\varepsilon$ is
related to variations of the wind direction in the vertical 
profile of turbulence and to its temporal variability 
during the exposure in a single layer. Numerical estimates 
(Lazorenko 2002b) show that even small $\pm 5 {\degr}$ variations 
in the wind direction cause an effect very similar to a complete averaging 
over 0--2$\pi$ range. The averaging leads to
expression (Lazorenko 2002a) 
 $ {\mathcal F}^2 \{ \mbox{rect} \}=(\pi V T q)^{-1}$ valid for 
long exposures $T \gg D/V$.
Finally, note that due to the forthcoming integration in the  
frequency plane, the angular-dependent function
${\mathcal F}^2 \{ \tilde{Q} \}$ in 
 Eq.(\ref{eq:2_9}) can be substituted by its average 
\begin{equation}
\label{eq:2_10a}
Q(q)=(2\pi)^{-1} \int_0^{2\pi}{\mathcal F}^2 \{
\tilde{Q}({\bf r,s} ) \}  d\varphi
\end{equation}
which is the filter induced by the reference star field.
Then Eq.(\ref{eq:2_9}) takes the form
\begin{equation}
\label{eq:2_10}
F_{\zeta}(q)= \frac{\lambda^2 c_\phi}{2\pi VT}
 q^{-2-p}Y(q)Q(q)
\end{equation}
where  $Y(q)={\mathcal F}^2 \{P/ \sigma \}$ is
the filter-function of the entrance pupil.
 Integration
of the power density (\ref{eq:2_10}) over $q$  
yields the variance of differential image motion
\begin{equation}
\label{eq:2_11}
\Delta ^2= \frac{\lambda^2 c_\phi}{VT} \int_{0}^{\infty} 
		Y(q)Q(q)q^{-1-p} \, dq 
\end{equation}

\section{Image motion power spectrum attenuation }
A concept of image motion variance  (\ref{eq:2_11}) reduction 
takes advantage of the dependence of the power spectrum  $F_{\zeta}(q)$
 on filter $Y(q)$ and $Q(q)$ shape which generally can be corrected.
To explain the concept of the method consider 
a classic example of double star observations
with a filled monopupil (\ref{eq:2_5}) when functions $Y(q)$ and 
$Q(q)$ are given by expressions  (e.g. Lazorenko 2002a)
\begin{equation}
\label{eq:2_12}
\begin{array}{l}
Y(q)=[2J_1(\pi Dq)/(\pi Dq)]^2\\
Q(q)=2[ 1-J_0 (2\pi qs_1)], \\
\end{array} 
\end{equation}
where $J_m$ are Bessel functions of the order $m$. 
In this study we suggest some improved modifications
of filters (\ref{eq:2_12})  for better filtration
of turbulent phase distortions. For this purpose we require that:
1) the low-pass filter $Y(q)$ is fast decreasing at frequencies
longer than $D^{-1}$, and 2) the high-pass filter $Q(q)$ has very low 
response
at short frequencies. A useful approximation of filter shapes is given
by quasi-rectangular functions with flat peaks and drop-down segments:
\begin{equation}
\label{eq:2_13}
\begin{array}{l}
Y(q)= \left \{ \begin{array}{ll}
               1,                       & \pi D q \leq q_{0} \\
           E_{\nu}(Dq/2)^{-\nu}, & \pi D q > q_{0}\\
               \end{array} \right.   
                                          \\
Q(q)= \left \{ \begin{array}{ll}
               H_{k}( Sq)^k,             & \pi S q \leq 1 \\
               Q_{\infty}, & \pi S q >1\\
               \end{array} \right.   
                                     \\
\end{array} 
\end{equation}
 Here the terms $\nu $ and $k$  are
the most important  
model  parameters since they determine the asymptotic  behaviour of 
filters; $E_{\nu}$, $H_{k}$ and $Q_{\infty}$ are  quasi-constant 
terms, approximately
independent of  $\nu $ and $k$;
 $q_{0}$ specifies the filter $Y(q)$ nucleus width; 
$S$ is the effective linear size of the stellar group  projection
onto the phase screen.  

In terms of this model, a very narrow field regime of
observations is defined by a nonoverlapping position of $Q(q)$ and 
$Y(q)$ nuclei: $S \ll D/q_0$. With an approximation $q_0 \approx 2$ 
following from Eq.(\ref{eq:3_3}) it is simply $S \ll D/2$.

Classic filters (\ref{eq:2_12}) are described by parameters
$\nu=3$ and $k=2$.

\begin{figure}[tbh]  %fig. 2
%\label{fig:qy}
\includegraphics*[ width=8.3cm, height=6.2cm]{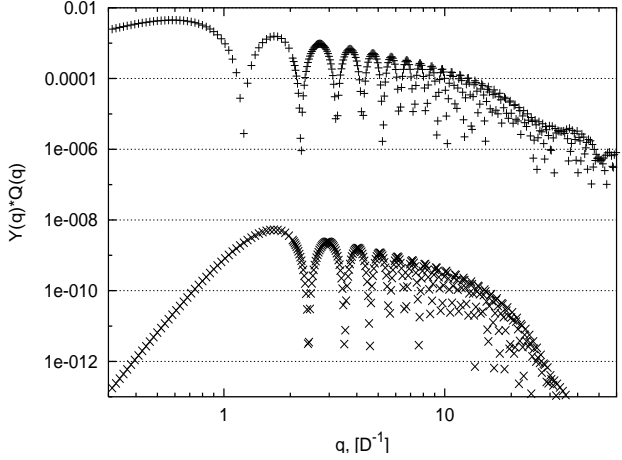}
\caption{Combined filter $Y(q)Q(q)$  response at $D/S=17.2$ for 
the two configurations of Table 2: "a" (upper curve) -- 
double star, $k=2$, $\nu=3$;  and  "g"  (lower curve) --
at $k=8$, $\nu=9$ }
\end{figure}

A qualitative demonstration of the current concept of enhanced image 
motion filtration requires using results from the next
Sections 4--5. In Fig.2 we display the combined filtering 
effect of the product $Y(q)Q(q)$ for a double star observation
($k=2$, $\nu=3$, upper curve) which should be compared with
that
expected for improved parameters $\nu=9$ and $k=8$ (lower curve). 
The functions plotted refer to a $D=100$ m
telescope, a single
turbulent layer at $h=20$ km and  star reference
groups "a"  and "g" of equal effective linear size $S=5.82$ m (Table 2).
The much lower  filter response in the second case ensures
the essential reduction of the differential image motion spectrum 
and  $\Delta^2$.

With the model (\ref{eq:2_13}), and the assumption of  
very narrow fields $S \ll D/2$, the value of $\Delta^2$ defined by 
the integral (\ref{eq:2_11}) is given by a two component sum 
\begin{equation}
\label{eq:2_14}
\Delta^2=[c_1(2S/D)^{k}+c_2(2S/D)^{\nu +p}]{(D/2)}^{p}{\pi}^{\nu -k +p}
\end{equation}
where $c_1$ and $c_2$ are some slow functions of the model parameters. 
Under condition $k<\nu +p$,
the first component prevails resulting in a power law  
$\Delta^2={(2S/D)}^kD^p$ while a dependence 
$\Delta^2=(2S/D)^{\nu +p}D^p$ is expected at $k>\nu +p$ . This situation 
is schematically reflected in Table 1 where the expected power
laws for $\Delta^2$ are shown. 
Diagonal elements of the table are marked with boxes and 
physically impossible $k$ (odd) are omitted.
\begin{table}[tbh]  %table 1
%\label{tab:asymp}
\caption[]{Asymptotic dependency of image motion variance $\Delta^2$
on $S$ and $D$}
\begin{flushleft}
\begin{tabular}{@{ }c|c@{\extracolsep{1mm}}c@{   }c@{   }c@{   }c@{   }c@{   }c@{ }} 
\hline
 &&&& $k$ & & \\
\cline{2-8}
  $\nu $ &       2    &   3&            4    & 5 &    6             & 7  & 8 \\
\hline
  3   & $S^2D^{-2+p}$ & \fbox{-}  & $S^{3+p}D^{-3}$ & - & $S^{3+p}D^{-3}$  &  - & 
        $S^{3+p}D^{-3}$  \rule{0pt}{11pt} \\
  4   & $S^2D^{-2+p}$ & -  & \framebox{$ S^{4}D^{-4+p}$} & - & $S^{4+p}D^{-4}$  &  - & 
        $S^{4+p}D^{-4}$ \\
  5   & $S^2D^{-2+p}$ & -  & $S^{4}D^{-4+p}$ & \framebox{-} & $S^{5+p}D^{-5}$  &  - & 
        $S^{5+p}D^{-5}$ \rule{0pt}{11pt} \\
  6   & $S^2D^{-2+p}$ & -  & $S^{4}D^{-4+p}$ & - & \fbox{$S^{6}D^{-6+p}$}  &  - & 
        $S^{6+p}D^{-6}$  \\
  7   & $S^2D^{-2+p}$ & -  & $S^{4}D^{-4+p}$ & - & $S^{6}D^{-6+p}$  &  \fbox{-} & 
        $S^{7+p}D^{-7}$  \rule{0pt}{11pt}\\
  8   & $S^2D^{-2+p}$ & -  & $S^{4}D^{-4+p}$ & - & $S^{6}D^{-6+p}$  &  - & 
        \fbox{$S^{8}D^{-8+p}$}  \\
  9   & $S^2D^{-2+p}$ & -  & $S^{4}D^{-4+p}$ & - & $S^{6}D^{-6+p}$  &  - & 
        $S^{8}D^{-8+p}$ \rule{0pt}{11pt} \\
\hline
\end{tabular}
\end{flushleft}
\end{table}
      
Remembering that very narrow field $2S/D \ll 1$ regime of observations 
is discussed, from Table 1 we find that the use of high 
$\nu$ and $k$ orders may result in a very strong reduction of $\Delta^2$ 
magnitude. Thus,
while power laws $\Delta^2 \sim D^{-2+p}$ and $\Delta^2 \sim S^2$
are valid at $k=2$ and $\nu=3$, application of
filters (\ref{eq:2_13}) with $\nu=9$ and $k=8$ 
leads to  much  stronger dependencies $\Delta^2 \sim D^{-8+p}$ and 
$\Delta^2 \sim S^8$, owing to which about a 256-fold change in
$\Delta^2$ magnitude is expected for a 2-fold change in $D$ or $S$.
In comparison to double star observations, 
the expected decrease in $\Delta^2$ is roughly  
$(2S/D)^6$.
For the case shown in Fig.2, the gain is
about ${(8.6)}^6 \approx 4 \cdot 10^5$. This estimate is rather
approximate; more reliable assessments are given by Eq.-s(\ref{eq:5_2})
and (\ref{eq:5_3}).

Table 1 shows that the optimal sets of $k$ and $\nu$ are 
concentrated near the diagonal, so the best 
suppression of image motion is expected at 
 $k \approx \nu$. A further increase of one of these parameters,
e.g. $k$ at fixed $\nu$, leads to no improvement in power law
though  it may essentially affect the magnitude of $\Delta^2$.

%% file: 1481_3.tex
\section{Apodization of the entrance pupil}
In this section we suggest an improvement of the filter
$Y(q)$ frequency response by applying a special apodizing
mask in the pupil plane. 
It is known that a proper apodization of the entrance pupil 
ensures  faster decay of the 
Fourier transform  ${\mathcal F}\{P\}$ of pupil function $P(r)$ 
at high 
frequencies (Papoulis 1971). In application to astrometric 
measurements, apodization allows one to suppress high-frequency 
phase distortions which is equivalent to  attaining
a high  $\nu$ order of the function $Y(q)$.

It is known (Papoulis 1971) that when the function $P(r)$ and 
its first $n-2$ derivates become zero at the pupil's edge $r=D/2$, then 
its  Fourier transform decreases asymptotically as $ q^{-n-1/2}$ 
at high $q$.
The function  $Y(q)={\mathcal F}^2 \{P\}$ in this case is
approximated by Eq.(\ref{eq:2_13}) with a parameter $\nu=2n+1$. 
To obtain high $\nu$ orders, it is therefore necessary to use
functions $P(r)$ which are sufficiently flat at  pupil edges.
Also, apodization should ensure sufficiently high 
light-transmission of the objective.
For instance, definition of $P(r)$ as a convolution of the two
functions (\ref{eq:2_5}), or $P(r)={\pi}^{-1}[2\arccos
(2r/D)-(4r/D)\sqrt{1-4r^2/D^2}]$ leads to $Y(q)=[2J_1(\pi Dq/2)/
(\pi Dq/2)]^4$ with $\nu=6$ and an effective collective 
area (\ref{eq:2_6}) of the objective 
$\sigma=\pi D^2/16$. A light-transmission of the objective
\begin{equation}
\label{eq:3_1}
\gamma=4 \sigma/(\pi D^2 )
\end{equation}
with this type of apodization is only $\gamma=1/4$. 
Better light-transmission is provided with the function 
\begin{equation}
\label{eq:3_2}
P(r)=(1-4r^2/D^2)^{n-1}, 
\end{equation}
where $n$ is an arbitrary positive integer. Functions $Y(q)$ 
correspondent to  (\ref{eq:3_2}) and
their characteristic parameters (\ref{eq:2_13})  are given below:
\begin{equation}
\label{eq:3_3}
\begin{array}{lll}
Y(q)=\left [ 2^n n! \frac{J_n(\pi Dq)}{(\pi Dq)^n} \right ]^2, &
\sigma=\frac{\pi D^2}{4n}, &  \gamma=n^{-1}, \\
 & & \\
E_{\nu}=(n!)^2/(2 \pi^{2n+2}), & \nu=2n+1, & q_0=\frac{2n}{\sqrt{2n-1}} \\
\end{array}
\end{equation}
An expression for effective filter width $q_0$ was found from the natural
condition $\int \int Y(\vec {q}) \, d \vec {q}= \pi {(q_0/ \pi D)}^2$;
at moderate $n$ $q_0 \approx 2$.

The functions $P(r)$ plotted for $n=1 \ldots 4$ and corresponding to 
$\nu=3 \ldots 9$ odd are shown in 
Fig.3. 
The filters  $Y(q)$ formed by this type of apodization are shown in 
Fig.4 and have asymptotes $ \sim q^{-\nu}$. 

\begin{figure}[tbh]  %fig. 3
%\label{fig:P}
\includegraphics*[ width=8.3cm, height=6.2cm]{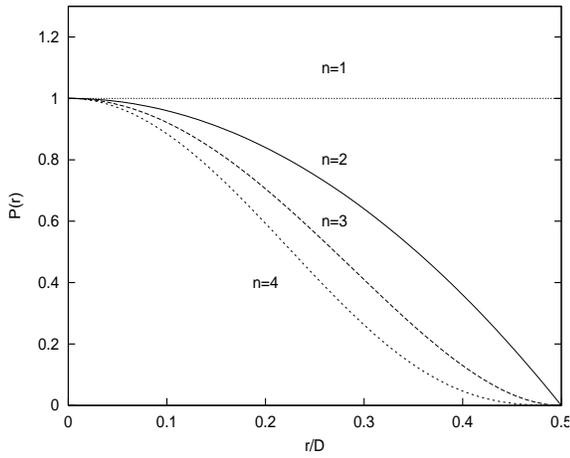}
\caption{The pupil functions $P(r)$ defined by Eq.(\ref{eq:3_2}) for
$n =1 \ldots  4 $ ($\nu =3 \ldots 9$)}
\end{figure}

\begin{figure}[tbh]  %fig. 4
%\label{fig:Y_q}
\includegraphics*[ width=8.3cm, height=6.2cm]{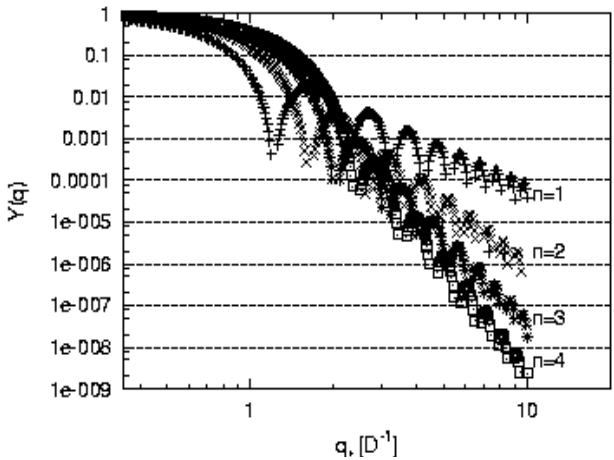}
\caption{Functions $Y(q)$ correspondent to $P(r)$ shown in 
Fig.3}
\end{figure}

A negative, inevitable consequence of high $\nu$ order implementation
is the decrease in light transmission $\gamma$ and so a weak light 
signal.
Thus a certain trade-off exists between good filtration of image 
motion and a high signal level that is necessary for image centroiding.

%% file: 1481_4.tex
\section{Reference frames with a virtual symmetry}
Consider a vector quantity $\vec {W}$ with two components
\begin{equation}
\label{eq:4_1}
\begin{array}{l}
W_x=  N^{-1} \sum_{i=1}^{N}a_i(\bar {x}_0-\bar {x}_i)=
	\bar {x}_0 - N^{-1} \sum_{i=1}^{N}a_i \bar {x}_i \\
W_y=  N^{-1} \sum_{i=1}^{N}a_i(\bar {y}_0-\bar {y}_i)=
	\bar {y}_0 - N^{-1} \sum_{i=1}^{N}a_i \bar {y}_i
\end{array}
\end{equation}
formed by a linear combination of measured (instant image motion included) 
differences of Cartesian coordinates of the target $\bar{x}_0$, $\bar{y}_0$ 
and $i$-th 
reference star $\bar{x}_i$, $\bar{y}_i$, $i=1,2..N$  scaled 
at the phase screen height $h$; 
$a_i$ are some weights that satisfy a normalizing condition 
\begin{equation}
\label{eq:4_1a}
\sum_{i=1}^{N} a_i=N.
\end{equation}

Introduction of the quantity (\ref{eq:4_1}) instead of
normally non-weighted  differential positions reduces the problem of
image motion suppression to 
the qualification of conditions on $a_i$ which minimize
the variance of $\vec {W}$.
A peculiarity of this study is the permission to use any, even
negative, weights $a_i$.
The vector  $\vec {W}$ that defines the differential position of the target 
object relative to the weighted center of reference
group $ N^{-1} \sum a_i \bar{x}_i$, $ N^{-1} \sum a_i \bar{y}_i$
is the only quantity which can be measured with high precision.
Therefore, the possibility to extract positional information from  $\vec {W}$ 
is not obvious and will be discussed in Section 7. 

Besides atmospheric noise, the total error of observations
depends also on the image-centroiding error component $\sigma_{ph}$ in 
$\vec {W}$ caused by a Poisson  photon noise  
in star images. Assuming that $D_i$ is the variance of the $i$-th reference 
star centroid position caused by  photon noise,  we find 
from Eq.(\ref{eq:4_1}) 
\begin{equation}
\label{eq:4_11}
\sigma^2_{ph}=  N^{-2}\sum_{i=1}^{N} a^2_i D_i
\end{equation}
since a contribution from the 
usually bright target object is small. 
For Gaussian shaped images with
r.m.s. width $\sigma_0=\mbox{FWHM}/2.36$ and $n_i$ detected
photons,  an expression 
(Irwin 1985) 
\begin{equation}
\label{eq:4_pp}
D_i=\sigma_0^2/n_i
\end{equation}
is valid for sufficiently bright images with a light signal exceeding
photon noise.

\subsection{Functions $Q(q)$ of high $k$-orders}
Expressions for 
differential image motion effects derived in Sections 2 -- 3
in application to the double star observations
can be easily extended to the case of a measured value
$\vec {W}$. Image motion for the quantity (\ref{eq:4_1}) is now
expressed through 
the weighted differences of the function
$\zeta$ values in points B$_0$ and B$_i$: 
$\Delta \zeta = N^{-1} \sum_i a_i[\zeta(B_0)-\zeta(B_i)]$.
 The difference operator 
(\ref{eq:2_8a})  corresponding to $\Delta \zeta$ and
the geometry shown in Fig.1 is then
\begin{equation}
\label{eq:4_3}
\tilde{Q}({\bf r,s})=N^{-1} \sum_i a_i[\delta ({\bf r})-\delta ({\bf r+s}_i)]
\end{equation}
A Fourier transform of this expression with a subsequent integral 
averaging (\ref{eq:2_10a}) yields a function 
\begin{equation}
\label{eq:4_4}
Q(q)= N^{-2} \sum_{i,j=1}^N a_i a_j [1-2J_0(2\pi q s_i)+
J_0(2\pi q s_{ij})]
\end{equation}
corresponding to a new definition of the measured quantity. Analysis
of Eq.(\ref{eq:4_4}) permits us to formulate the
conditions necessary to increase the  $Q(q)$ function  order.
For this purpose we expand the Bessel functions in (\ref{eq:4_4}) into 
power series  of $q$, deriving the approximation at
low frequencies:
\begin{equation}
\label{eq:4_5}
\begin{array}{ll}
Q(q) & = \frac{1}{N^{2}} \sum_{m=1}^{\infty} \frac{(-1)^{m+1}}{(m!)^2}
	(\pi q)^{2m}
	 \sum_{i,j=1}^{N} a_i a_j \\
 & \\
     & \times (s_i^{2m}+s_j^{2m}- s_{ij}^{2m})
\end{array}
\end{equation}
The order $k$ of the $Q(q)$ function is determined by the first
non-zero term in the expansion; normally, $k=2$ if no special
efforts are made. 
 
With $s_i$, $s_{ij}$ fixed, one may always  choose
a set of coefficients $a_i$  turning to zero the first or few  first
 sum values $\sum_{i,j=1}^{N} a_i a_j (s_i^{2m}+s_j^{2m}- 
s_{ij}^{2m})$ which are coefficients at $q^{2m}$. With
a sufficiently large number of reference stars, all the first 
$q^2, q^4 \ldots q^{k-2}$ expansion terms in (\ref{eq:4_5}) could be 
eliminated by applying conditions
\begin{equation}
\label{eq:4_6}
\begin{array}{l}
  \sum \sum a_i a_j (s_i^{2}+s_j^{2}- s_{ij}^{2})=0,\\
  \sum \sum a_i a_j (s_i^{4}+s_j^{4}- s_{ij}^{4})=0, \\
  \ldots                                         \\
  \sum \sum a_i a_j (s_i^{k-2}+s_j^{k-2}- s_{ij}^{k-2})=0
\end{array}
\end{equation}
where $k $ is some even integer.
Under conditions (\ref{eq:4_6}), the power of the first non-zero 
term in the expansion (\ref{eq:4_5}) can be increased 
from normal $k=2$ to some higher $k \ge 4$.
Eq.-s(\ref{eq:4_6}) is a system with 
unknowns $a_i$; 
it includes only a single first line at $k=4$, the two
first lines at $k=6$, and so on. The system (\ref{eq:4_6}) can be 
converted to a simpler form. Thus, passing to
Cartesian coordinates $x_i-x_0=s_i\cos\psi _i$, 
$y_i-y_0=s_i\sin\psi _i$ and
taking into consideration   Eq.(\ref{eq:4_1a}) one finds that each sum of
sums in Eq.-s(\ref{eq:4_6}) takes the form of products of one-dimensional 
sums.
For instance, the first equation leading to $k=4$ transformes to
\begin{equation}
\label{eq:M_x}
 \sum \sum a_i a_j(s_i^{2}+s_j^{2}- s_{ij}^{2})=2(M_x^2+M_y^2)=0
\end{equation}
where $M_x=\sum a_i(x_i-x_0)$ and $M_y=\sum a_i(y_i-y_0)$ are the 
first coordinate moments. The next equation, for $k=6$,  
$ \sum \sum a_i a_j(s_i^{4}+s_j^{4}- s_{ij}^{4})=2[\sum a_i(x_i-x_0)^2+
\sum a_i(y_i-y_0)^2]^2+ 4[\sum a_i(x_i-x_0)^2]^2 +
4[\sum a_i(y_i-y_0)^2]^2 +8[\sum a_i(x_i-x_0)(y_i-y_0)]^2=0$ contains
quadratic cross-moments.
Direct computations show that the order of cross-moments is incremented 
by 1 when passing
to each next $k$ order. The quadratic form of the equations suggests that all
weighted cross moments of reference star coordinates are zero. 
 An equivalent form of system (\ref{eq:4_6}) is  therefore
\begin{equation}
\label{eq:4_7}
\begin{array}{l}
\sum a_i =N,                             \\
\sum a_i (x_i-x_0)=\sum a_i (y_i-y_0)=0,                \\
\sum a_i (x_i-x_0)^2=\sum a_i (x_i-x_0) (y_i-y_0)=\\
=\sum a_i (y_i-y_0)^2=0,  \\
\ldots  \\
\sum a_i (x_i-x_0)^{\frac{k}{2}-1}=\sum a_i (x_i-x_0)^{\frac{k}{2}-2}
      (y_i-y_0)=\\
\ldots = \\

	 \sum a_i (x_i-x_0)(y_i-y_0)^{\frac{k}{2}-2}= 
	\sum a_i (y_i-y_0)^{\frac{k}{2}-1}=0  
\end{array}
\end{equation}
which reveals the modal structure of the filter $Q(q)$. 
At $k=2$, when none of conditions (\ref{eq:4_6}) are fulfilled, 
the system is limited by a single first line normalizing equation; 
at $k=4$ it includes the first two lines (three equations), 
and so on; the total
number of equations is $k(k+2)/8$. In a compact form, the
system (\ref{eq:4_7}) with unknowns $a_i$ is written as
\begin{equation}
\label{eq:4_7a}
\begin{array}{lr}
\sum a_i =N,       & \\
\sum a_i (x_i-x_0)^{\alpha} (y_i-y_0)^{\beta}=0,& \; 
		\alpha + \beta = 1 \ldots  \frac{k}{2}-1,   
\end{array}
\end{equation}
where $\alpha$ and $\beta$ are positive integers. The use of 
weights $a_i$ satisfying  
Eq.-s(\ref{eq:4_7a}) results in elimination of $k/2-1$ first  terms 
in the expansion (\ref{eq:4_5}) which becomes
\begin{equation}
\label{eq:4_8}
\begin{array}{ll}
Q(q)  & = \frac{1}{N^{2}} \sum_{m=k/2}^{\infty} \frac{(-1)^{m+1}}{(m!)^2}q^{2m}
	 \sum_{i,j=1}^{N} a_i a_j \\
  & \\
     &  \times (s_i^{2m}+s_j^{2m}- s_{ij}^{2m})
\end{array}
\end{equation}
It should be noted that in the case of $k \geq 6$ when quadratic moments
of $x$ and $y$ are involved in  Eq.-s(\ref{eq:4_7a}),
the solution  must incorporate negative $a_i$ values, which is 
rather unusual for a common technique of astrometric reductions. 

Application of various sets of $a_i$ changes the 
geometric properties of a reference group. Thus, implementation
of $a_i$ of
the $k=4$ order for which condition $\sum a_i(x_i-x_0)=
\sum a_i(y_i-y_0)=0$
is fulfilled  transforms an arbitrary
reference group into its virtual equivalent with an ideal
symmetric
structure  centered at the point $x_0$, $y_0$. 
With respect to the image motion
statistics, and the filter $Q(q)$ asymptotic behaviour in particular, 
both groups 
become indistinguishable providing they are of equal effective size.
A further increase of $k$ results in enhanced improvement
of reference group filtering properties which can reveal  
a power dependency stronger than $Q(q) \sim q^4$, a case 
impossible with a simple geometric symmetry. 

Implementation of high $k$ orders   is
limited by the number $N$ of reference stars  available. 
The minimum $N$ value required to achieve some $k$ order and to find 
solutions $a_i$ of the  system (\ref{eq:4_7a}) is 
\begin{equation}
\label{eq:4_9}
N_{min}= \left \{
  \begin{array}{lr}
    k(k+2)/8, & \mbox{for 2-D distribution of stars}\\
    k/2,   &    \mbox{for 1-D, in-line distribution}   
  \end{array}   \right.
\end{equation}
Here we admitted that $N_{min}$, generally, depends on the type of star 
distribution in the field. Though exactly linear configurations 
do not exist, they serve well for illustrative 
purposes. 

At $N>N_{min}$, 
a redundancy  of the system (\ref{eq:4_7a}) allows us to set a useful 
(even in the case of $k=2$) condition   
\begin{equation}
\label{eq:4_10}
\sum a^2_i D_i =\mbox {min}
\end{equation}
which  reduces the centroiding error (\ref{eq:4_11}). A consistent 
solution of (\ref{eq:4_7a}) and (\ref{eq:4_10}) is found 
with  a standard Lagrangian method of undetermined 
coefficients. Such a solution is optimal with respect to both atmospheric
and photon centroiding errors.

\subsection{Tutorial linear configurations}
Some tutorial examples of linear (along the $x$-axis) 
configurations are given in
Table 2 which contains the conditional name of configurations,
 $x$-coordinates 
of stars given with reference to the target and expressed in seconds 
of arc, $a_i$ values, $\overline{ a^2}=N^{-1}\sum a_i^2$ and
the principal $q^k$  term of the $Q(q)$ function expansion. $x_i$ are 
scaled  to equalize the effective angular 
size of each group to $\rho=1'$. For a turbulent layer at $h=20$ km, 
this angle corresponds to an effective linear size $S=\rho h=5.82$ m. 
The last quantity is convenient to define as
\begin{equation}
\label{eq:4_12}
S= \left | N^{-2} \sum_{ij}^N a_i a_j (s_i^k+s_j^k-s^k_{ij}) \right |^{1/k}
\end{equation}
which relates $S$ to the magnitude of the first non-zero term in the 
expansion (\ref{eq:4_5}). 
A convenience of this definition  is that 
series expansion (\ref{eq:4_8}) (its leading term) is now
simplified to 
\begin{equation}
\label{eq:4_13}
Q(q)= (\pi q S)^k /[(k/2)!]^2
\end{equation}
Note that effective sizes $S$ and $\rho$  depend
very weakly on the peculiarities of the star distribution in the frame and 
approximately are equal or slightly exceed its radius (the largest 
separation target-reference star).

\begin{table}[tbh]  %table 2
%\label{tab:3}
\caption[]{Linear configurations of effective angular size $\rho=1'$ 
and $k$-order virtual symmetry}
\begin{flushleft}
\begin{tabular}{@{ }l@{ }lllll@{ }l@{ }} 
\hline
 Name  &  $N$  & $k$ & $x_i$ & $a_i$ & $\overline{ a^2}$ & $Q(q)$ \rule{0pt}{11pt}\\
\hline
a  & 1 & 2 & 42$''$  &    1   &       1   &   $q^2$  \rule{0pt}{11pt} \\
b  & 2 & 4 & -38, 38  &    1, 1   &     1   &   $q^4$  \\
c  & 2 & 4 & 27, 54  &    4, -2   &     10   &   $q^4$  \\
d  & 3 & 6 & -29, 29, 58  &    1, 3, -1   &   3.67   &   $q^6$  \\
e  & 3 & 4 & -32, 32, 64  & 
    12/7,6/7,3/7   &   9   &   $q^4$  \\
f  & 4 & 4 & -31$\sqrt{2}$, -31,  &     &&\\
	&&&		31, 31$\sqrt{2}$  
			&   1, 1, 1, 1   &       1   &   $q^4$  \\
g  & 4 & 8 & -30$\sqrt{2}$, -30,  &&&\\
	&&&		30, 30$\sqrt{2}$  
			 &   -2, 4, 4, -2   &      10   &   $q^8$  \\
h  & 6 & 12 & -56,-37.4,-18.7, & 0.3,-1.8,4.5, &&\\
       &&&     18.7,37.4,56 & 4.5,-1.8,0.3   &   7.9   &   $q^{12}$  \\
\hline
\end{tabular}
\end{flushleft}
\end{table}

Table 2 shows that even strongly asymmetric star groups 
"c" and "d" are
subject to high order symmetrization. This is achieved, however, at the
expense of applying large $a_i$ owing to which the $\overline{ a^2}$ value
is greater than that for symmetric groups (compare cases "b" and "c").
Thus, application of high order 
symmetry for asymmetric star groups reduces atmospheric noise
but causes a rise of the centroiding error (\ref{eq:4_11}). The same
effect is observed even for symmetric distributions "g" and "h" when 
very high $k$ is used. 
\begin{figure}[tbh]  %fig. 5
%\label{fig:Q_q}
\includegraphics*[ width=8.3cm, height=6.8cm]{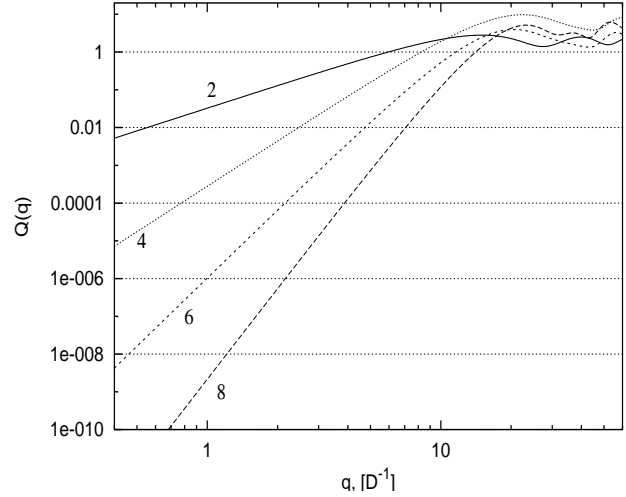}
\caption{Functions $Q(q)$ for stellar groups "a, c, d, g"
(Table 2) with $k=2$, 4, 6 and 8 symmetry order; $D/S=17.2$ }
\end{figure}

The $Q(q)$  plots shown in  Fig.5 for a few configurations of Table 2
emphasize a difference in the function form at low 
frequencies  which for
both symmetric and non-symmetric star distributions is determined 
only by the index $k$. 
Asymmetry of star groups
 "c" and "d" leads to the increase of amplitude
$Q_{\infty}$ of $Q(q)$ at high frequencies $q \ge 1/S$, 
which, due to relation
$\overline{ a^2}/N= Q_{\infty}-1$ following from Eq(\ref{eq:4_4}), is 
indicative of the $\sigma^2_{ph}$ increase.

An example of 2-D stellar group symmetrization
is given in Table 3 for a  sample  of  $N=14$ stars of the
open cluster NGC 2420 observed with the 5-m 
Palomar telescope (Pravdo \& Shaklan 1996).
The table contains star coordinates $x_i$, $y_i$ with reference to
the target star, weights $a_i$ and $\overline{ a^2}=\sum a_i^2/N$ 
for various orders  up to $k=8$. Weights $a_i$ were computed using
conditions (\ref{eq:4_10}) with $D_i=\mbox {const}$.
The extremally small angle $\rho=3.7''$ found at $k=2$ is by no
means due to a special selection of reference stars (no selection 
was applied) and reflects the effect of field averaging
making the first moments $M_x$ and $M_y$ small for large $N$.
In the limit of $N \to \infty$, discussed in Section 6, any $k=2$
order group is moved up to the 4-order.
Note also that the value of $\overline{ a^2}$ is
rapidly increasing with $k$, so does the variance  (\ref{eq:4_11}).

\begin{table}[tbh]  %table 3
\caption[]{ Weights $a_i$ symmetrizing a stellar group 
(Pravdo \& Shaklan 1996) to $k=2 \ldots 8$ orders }
\begin{flushleft}
\begin{tabular}{@{ }rr|rrrr@{ }} 
\hline
        &&& \quad $ k$ &  &\\
\cline{3-6}
  $x_i$  &  $y_i$    & 2 & 4 & 6 & 8 \\
\hline

   46.0$''$ &  19.4$''$  & 1.0   &  0.914   & -1.355   &  -0.023  \\
   33.5 &  17.1  & 1.0   &  0.908   &  0.340   &  -1.120  \\
   19.1 &  18.7  & 1.0   &  0.845   &  0.739   &   0.680  \\
    9.5 &  16.1  & 1.0   &  0.851   &  1.792   &   0.308  \\
   -3.2 &   5.7  & 1.0   &  0.951   &  3.414   &   3.300  \\ 
   17.1 &   3.4  & 1.0   &  1.039   &  3.201   &   4.300  \\ 
    8.8 & -11.4  & 1.0   &  1.209   &  1.787   &   7.102  \\ 
   14.6 &  -15.7 &  1.0  &   1.282  &   0.624  &   -0.265 \\  
   19.2 &  -16.1 &  1.0  &   1.300  &   0.465  &   -2.793  \\ 
  -24.1 &    1.1 &  1.0  &   0.950  &   2.981  &    2.157  \\ 
  -32.5 &   -4.7 &  1.0  &   1.002  &   1.940  &    2.362  \\ 
  -41.3 &   -8.6 &  1.0  &   1.028  &   0.572  &   -1.593  \\ 
  -36.3 &  -16.4 &  1.0  &   1.144  &  -1.407  &   -0.138  \\ 
  -45.5 &   24.9 &  1.0  &   0.578  &  -1.092  &   -0.278  \\ 
\hline
$\overline{ a^2}$ &  & 1.0  &  1.03  &  3.40  & 7.32  \rule{0pt}{11pt} \\       
\hline                                                    
$\rho$        &    & 3.7$''$ & 45.7$''$  & 35.2$''$   &   42.2$''$ 
			\rule{0pt}{11pt} \\       
\hline                                                    
\end{tabular}
\end{flushleft}
\end{table}

\subsection{Expressions for the variance of image motion}
Analytic expressions for $\Delta^2$ in a limiting case of very 
long exposures and narrow fields $T \gg D/V \gg S/V$ 
 can be derived from  
Eq.(\ref{eq:2_11}) where $Y(q)$  and
$Q(q)$  are given by Eq.-s(\ref{eq:3_3}) and (\ref{eq:4_4}):
\begin{equation}
\label{eq:5_1}
\begin{array}{ll}
\Delta^2  & =\frac{\lambda^2 c_{\phi}2^{\nu-1}
\left ( \frac{\nu-1}{2}! \right )^2}{N^2 VT} \int \limits_0^{\infty}
	 \frac{J_{(\nu-1)/2}^2 (\pi Dq)}
	{(\pi Dq)^{\nu-1}} 
  \sum \sum a_i a_j [1-  \\
  & -2J_0(2\pi q s_i)+J_0(2\pi q s_{ij})] q^{-1-p} \, dq
\end{array}
\end{equation}
An approach to integration  depends on the ratio  
between the  
$k$ and $\nu$ values. When $k>\nu +p$, the integral converges at $q=0$
so  approximation  (\ref{eq:2_13}) for $Y(q)$ is valid.
Integrating by parts and taking advantage of the condition
(\ref{eq:4_6}) results in
\begin{equation}
\label{eq:5_2}
\begin{array}{l}
\Delta^2  =\frac{\lambda^2 c_{\phi}2^{\nu}
\left ( \frac{\nu-1}{2}! \right )^2 \hat{ S}^{\nu+p}}
       {4{\pi}^{1-p}VT D^{\nu}{\Gamma}^2(\frac{p+\nu+2}{2})}
  \left \{  \begin{array}{ll}
 \Gamma(\frac{p+1}{2})\Gamma(\frac{1-p}{2}), & \nu \mbox{- odd}\\
  \Gamma(\frac{p}{2})\Gamma(\frac{2-p}{2}), & \nu \mbox{- even}\\
                 \end{array}          \right.
	\\
  k>\nu +p \\
\end{array}
\end{equation}
Here $ \hat{S} ={|N^{-2}\sum \sum a_i a_j [2{s_i}^{p+\nu}-
{s_{ij}}^{p+\nu}]|}^{1/(p+\nu)} $ is a modified field size defined 
similarly to (\ref{eq:4_12})  but  with fractional powers. 
Since $s_i=s \rho_i /\rho$, from (\ref{eq:5_2}) we find
a power dependency $\Delta^2 \sim S^{\nu+p}/D^{\nu}$   
which is in agreement with Eq.(\ref{eq:2_14}) 
 and  Table 1  testifying that
an increase of $k$ over $\nu +p$ does not affect the power laws. 
Another interesting point is the dependency of $\Delta$ on $k$ which, 
however, is not  clear from (\ref{eq:4_6})
since it contains the $ \hat{S}$ term related to $k$ via  weights $a_i$.
A rough expression for $\Delta$ with no $ \hat{S}$ term can be 
obtained based on the
next Sections results  for non-discrete star fields.
Using  approximation (\ref{eq:6_8}) for $Q(q)$ and recomputing the 
integral (\ref{eq:2_11})  with $Y(q)$ given 
by (\ref{eq:2_13}), yields
\begin{equation}
\label{eq:5_ap1}
\begin{array}{ll}
\Delta^2  =\frac{\lambda^2 c_{\phi}2^{\nu-1}{(k/4)}^{k-\nu-p}
		 {(\frac{\nu-1}{2}!)}^2 {S}^{\nu+p}}
       { VT {[(k/2)!]}^2 (k-\nu-p) D^{\nu}} 
  &,  k>\nu +1 \\
\end{array}
\end{equation}
which is a decreasing function of $k$ at $\nu$ fixed.
The use of high $k$ orders, thus, always reduces atmospheric noise.

 At $k<\nu +p$, the integral (\ref{eq:5_1}) is calculated with
approximation (\ref{eq:4_13}) for $Q(q)$. Direct integration yields
\begin{equation}
\label{eq:5_3}
\begin{array}{l}
\Delta^2  =\frac{\lambda^2 c_{\phi}2^{\nu-2}{\pi}^p {(\frac{\nu-1}{2}!)}^2
		 \Gamma(\frac{k-p}{2})\Gamma(\frac{\nu+p-k}{2}) S^k}
       { VT {[(k/2)!]}^2 \sqrt{\pi}D^{k-p}\Gamma(\nu+\frac{p-k}{2})
        \Gamma(\frac{\nu+p+1-k}{2})} \\
  k<\nu +p \\
\end{array}
\end{equation}
and is also consistent with Table 1. It should be stressed that,
unlike the previous case of $  k>\nu +p$, an increase of  $\nu$
over $k$ gives rise to $\Delta^2$. The deterioration of the
results is related to the expansion of the filter $Y(q)$ nucleus width,
clearly seen in Fig.4.

%% file: 1481_5.tex
\section{Virtual symmetry for dense reference frames}
Approximate characteristics of reference groups with 
several stars
are easily found in a limit of infinite $N$ when the star distribution 
becomes
continuous. All estimates are found especially easily since
discrete summations are substituted by integrals, and
individual features of star distribution in the field become unimportant. 

\subsection{Approximate expressions for  $a_i$ and $\sigma_{ph}$ }
We assume equal brightness of stars. Assuming also that 
reference stars
are evenly distributed around the target in a circle 
of radius $R$ with a spatial
density $N/(\pi R^2)$, we introduce, instead of weights $a_i$,
a weighting function $a(r)$ with a radial symmetry which satisfies
the normalizing condition $\frac{N}{\pi R^2}\int_0^R a(r) \, dx \, dy=N$ 
equivalent to (\ref{eq:4_1a}). For a $k$-order function $Q(q)$, 
an integral analogue of the system of equations (\ref{eq:4_7a}) with
unknown function $a(r)$ is 
\begin{equation}
\label{eq:6_1}
\begin{array}{ll}
\int_0^R a(r)r \, dr  = R^2 /2, & k=4 \\
\int_0^R a(r)r^{\alpha} \, dr  = 0, & \alpha= 1,3 \ldots \frac{k-2}{2},
			k=8,12 \ldots \\
\end{array}
\end{equation}
The above system is valid for $k$ multiples of 4 only, since
for any symmetric distributions 
Eq.-s (\ref{eq:4_7a}) with coordinate cross-moments of odd powers are 
satisfied automatically; 
for this reason orders $k=2,6,10 \ldots $ do not exist, they are
moved up into the next higher order. Considering 
only a polynomial class of solutions for $a(r)$
of the form $a(r)=b_0 + b_2 r^2 + \ldots + 
b_\beta r^{\beta}$, where $\beta= k/2 -2$, and
performing integration  (\ref{eq:6_1}), we come to
a linear system of $\beta /2 +1$ equations with respect to unknowns 
$b_{2j}$:
\begin{equation}
\label{eq:6_2}
\sum _{j=0}^{\beta /2} \frac{b_{2j}R^{2j}}{2(i+j+1)}= \left \{
               \begin{array}{l}
		 R^2, \; i=0\\
		 0, \; i=1,2 \ldots \beta /2
		\end{array}
	\right.
\end{equation}
The solution found by computer simulation is
\begin{equation}
\label{eq:6_3}
\begin{array}{l}
a(r)=\frac{k}{4} \sum _{i=1}^{k/4}(-1)^{i+1}\frac{(k/4-i+1)!}{(k/4-i)!i!(i-1)!}
			\left ( \frac{r}{R} \right ) ^{2i-2},\\
		k=4,8 \ldots; \;  \; r \leq R 			
\end{array}
\end{equation}

\begin{figure}[tbh]  %fig. 6
%\label{fig:geom}
\includegraphics*[ width=8.3cm, height=6.2cm]{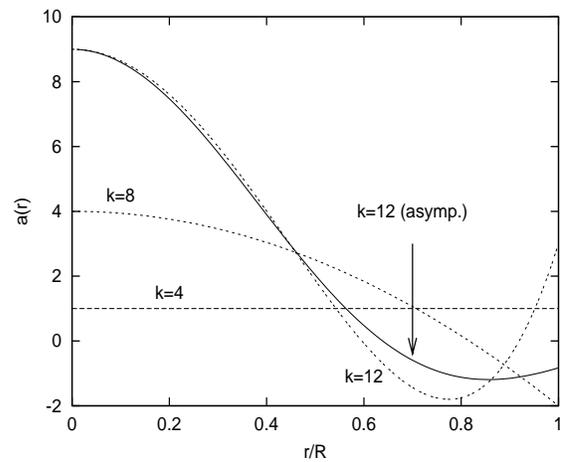}
\caption{  Weighting functions (\ref{eq:6_4}) for circular 
star distributions at $k=4$, 8, 12;  the
asymptotic form  (\ref{eq:6_5}) of $a(r)$ at $k=12$ }
\end{figure}

For a few first $k$, the weighting functions 
\begin{equation}
\label{eq:6_4}
a(r)= \left \{   \begin{array}{ll}
		1, & k=4\\
		2(2-3r^2/R^2), & k=8\\
                3(3-12r^2/R^2+10r^4/R^4), & k=12
\end{array}
		\right.
\end{equation}
are shown in Fig.6. Except for the case of $k=4$ discussed  by
Lindegren (1980) and when $a(r)=$const, the plots of $a(r)$ incorporate
some oscillating circular zones coming from the field outer borders.
The number of these zones and the function's peak increases  
with $k$ while the  amplitude of oscillations
decreases. These features are easily explained since 
Eq.(\ref{eq:6_3}) at $k \to \infty$ is reduced to 
$a(r)=(k/4)^2\sum_{i=0}^{k/4-1}(-1)^i
[kr/(4R)]^{2i}/[i!(i+1)!]$ which is a truncated power expansion of the 
Airy-type oscillating function
\begin{equation}
\label{eq:6_5}
a(r)=\left ( \frac{k}{4} \right )^2 \frac{2J_1(\frac{kr}{2R})}{kr/(2R)}
\end{equation}
For comparison, the asymptotic and exact forms of $a(r)$ for $k=12$ are 
shown in Fig.6.

Above we assumed equal brightness of field stars. Let us find 
expressions for a weighting function $a(r,D_i)$ that approximates
weights $a_i$ for descrete very dense distributions while 
accounting for 
magnitude-dependent centroiding variances $D_i$ (\ref{eq:4_pp}). 
In this case the condition (\ref{eq:4_10})  leads to
the inverse proportion $a_i \sim 1/D_i$, which follows also from
a least-squares principle. 
Assuming that spatial and brightness distribution of stars 
in the field are uncorrelated and using $a(r)$ as approximation
for $a_i$ at $D_i=$const, we obtain
\begin{equation}
\label{eq:6_p1}
a(r,D_i) \simeq a(r)\frac{D_1}{D_i}
\end{equation}
where $D_1 =N/\sum_{i=1}^N D_i^{-1}$ is the mean (effective) variance of 
one reference star. The total
centroiding variance of all reference field than is
$D_{\rm{eff}}=D_1 /N= 1/\sum D_i^{-1}$ . Taking Eq.(\ref{eq:4_pp}) 
into consideration, we find
\begin{equation}
\label{eq:6_p2}
 D_{\rm{eff}}=\sigma_0^2/ \sum n_i.
\end{equation}
Incorporating
fainter stars, a value of $ D_{\rm{eff}}$ always decreases (improves)  
but rapidly approaches some limit.
\begin{figure}[tbh]  %fig. 7
\includegraphics*[ width=8.3cm, height=6.2cm]{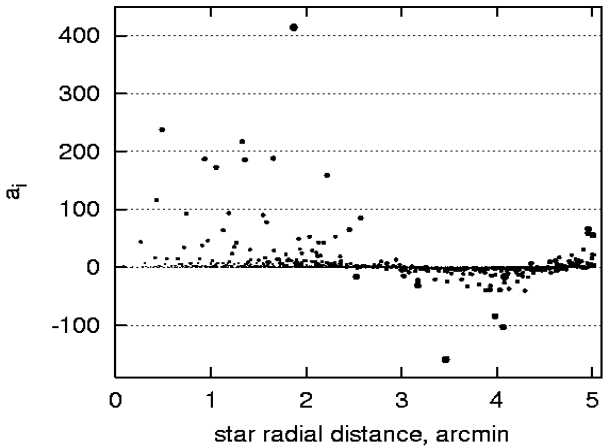}
\caption{  An example of $a_i$ coefficients symmetrizing
a random $5'$ field with 3170 stars to the $k=12$ order; dot sizes
are proportional to the star brightness (12 to 23 mag)}
\end{figure}

The above considerations are illustrated by a numerical simulation 
of a random 5$'$ field
(radius) with 3170 stars.  The magnitude  and sky star 
distribution corresponds to the Galaxy model 
(Bahcall \& Soneira 1980) and galactic
coordinates $b=20^0$, $l=0^0$. 
The field with stars to V=23 mag was symmetrized to the $k=12$ order;
a distribution of computed $a_i$ values versus distance from the 
field center is shown in Fig.7. In this example an accumulated
value of $D_{\rm{eff}}$ is equivalent to that produced by 120.3 stars
of 15 mag; an effective mean centroiding variance of one reference star 
$D_{1}$ corresponds to the star of V=18.5 mag. A large scatter of
computed $a_i$ values seen in Fig.7 is natural since $a_i \simeq 1/D_i$;
this relation is exact at $k=2$ and becomes rough at large $k$.
Usually, $|a_i| \gg 1$ for  bright and $|a_i| \ll 1$ for faint images. 
Sampled values of $a_i D_i /D_{1}$, on the contrary, show 
small scatter and a strong concentration around the  
$a(r)$ function  plot (Fig.8).
Approximate values of $a_i$ thus can be found from
Eq.(\ref{eq:6_p1}) for any $k$ star.
\begin{figure}[tbh]  %fig. 8
\includegraphics*[ width=8.3cm, height=6.2cm]{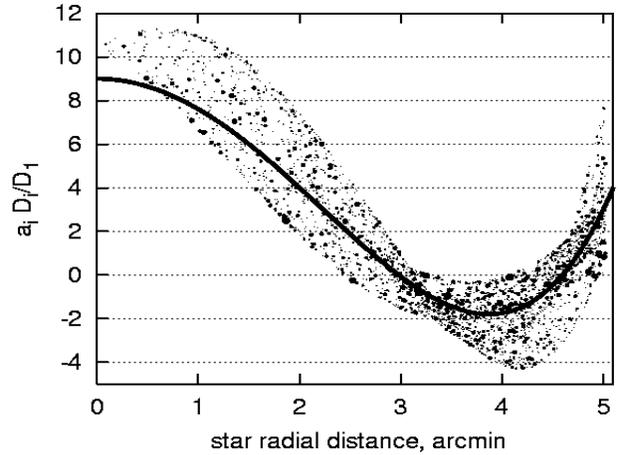}
\caption{   $a_i D_i /D_{1}$ values for
random data shown in Fig.7; solid line - approximation by the function
$a(r)$ plotted for $k=12$ }
\end{figure}

An increase of the $a(r)$ amplitude  in the field center
for high $k$ leads to an
increase of the total centroiding error $\sigma^2_{ph}$. 
This effect is easily estimated since, with above the
assumptions, the magnitude-related $D_i$ and
coordinate-related  $a_i^2$  terms in Eq.(\ref{eq:4_11}) are 
statistically independent. Therefore Eq.(\ref{eq:4_11}) 
transforms to $\sigma^2_{ph}=D_{\rm{eff}}  \overline{a^2}$ where 
$\overline{a^2}=(\pi R^2)^{-1}\int \int a^2(r) \, dx \,dy$  is  
an averaged value of $a(r)$. Direct integration of  
Eq.(\ref{eq:6_4}) yields $\overline{a^2}=(k/4)^2$ whence
\begin{equation}
\label{eq:6_6}
\sigma^2_{ph}=D_{\rm{eff}} (k/4)^2 
\end{equation}
The last equation can be given in terms of star image parameters. 
Taking into consideration  Eq.(\ref{eq:4_pp}) and 
 allowing for light transmission
$\gamma$ of the apodized objective  Eq.-s(\ref{eq:3_3}), we obtain
\begin{equation}
\label{eq:6_6b}
\sigma_{ph}=\frac{\mbox{FWHM}}{2.36 \sqrt{\sum n_i}} (k/4) 
	\sqrt{\frac{\nu -1}{2}}
\end{equation}
The photon noise thus depends on
FWHM, the total light of reference stars  $\sum n_i$, $k$ and 
$\nu $. The use of very faint stars as a reference clearly does not
lead to improvement in $\sigma_{ph}$ because their contribution
to $\sum n_i$ is negligible; 
the use of very high $k$ orders degrades $\sigma_{ph}$.

\subsection{Expressions for $Q(q)$ and $S$}
The asymptotic form of the function $Q(q)$ for stars of equal
brightness is found from a limiting
($N \to \infty$ ) form of the function (\ref{eq:4_3}) which  
is 
$\tilde{Q}({\bf r})=\delta ({\bf r})-P({\bf r}) a({\bf r})/(\pi R^2)$.
From this it follows that 
$Q(q)={\mathcal F}^2\{ \tilde{Q} \}=[1-2R^{-2}\int_0^R a(r)
J_0(2\pi r q)r \, dr]^2$ , and a direct integration with $a(r)$ given
by Eq.(\ref{eq:6_4}) yields
\begin{equation}
\label{eq:6_7}
Q(q)= \left \{ \begin{array}{ll}
               [1-2J_1(2 \pi Rq)/(2 \pi Rq)]^2, & k=4 \\
     \left [ 1 + \frac{4J_1(2 \pi Rq)}{(2 \pi Rq)}
	-\frac{24J_2(2 \pi Rq)}{(2 \pi Rq)^2} \right ] ^2, 
			& k=8 \\
& \\
\left  [ 1 - \frac{6J_1(2 \pi Rq)}{(2 \pi Rq)}
	+\frac{96J_2(2 \pi Rq)}{(2 \pi Rq)^2}- \right. & \\
\left.	-\frac{480J_3(2 \pi Rq)}{(2 \pi Rq)^3}  \right ] ^2, 
			& k=12 \\
 \ldots &\\
        
        \left \{
	\begin{array}{ll}
         0,  & q<k/(4  \pi R)\\
	 1,  & q>k/(4  \pi R)
	\end{array}     \right.  & k \to \infty \\
         \end{array}          \right.
\end{equation}
An expression for $k \to \infty$ derived with a Hankel transform
of Eq.(\ref{eq:6_5}) shows that an asymptotic form of $Q(q)$ is
an opaque  circle with radius proportional to $k$.
Expansion of Eq.-s(\ref{eq:6_7}) into power series of $q$  yields
an expression
\begin{equation}
\label{eq:6_8}
Q(q)=  \left \{ \begin{array}{ll}
              	\frac{(\pi Rq)^k}{[(k/2)!]^2}, &  q<k/(4  \pi R)\\
              	1, &      q>k/(4  \pi R)
	       \end{array}    \right.
\end{equation}
valid for $k=4,8 \ldots$. Comparing this expression
with a definition  (\ref{eq:4_13}) for effective size $S$, we come to
a very simple relation
\begin{equation}
\label{eq:6_9}
S=     \begin{array}{lrr}
              	R, &  k/2 & \mbox{even}\\
       \end{array}   
\end{equation}

\begin{figure}[tbh]  %fig. 9
\includegraphics*[ width=8.3cm, height=6.2cm]{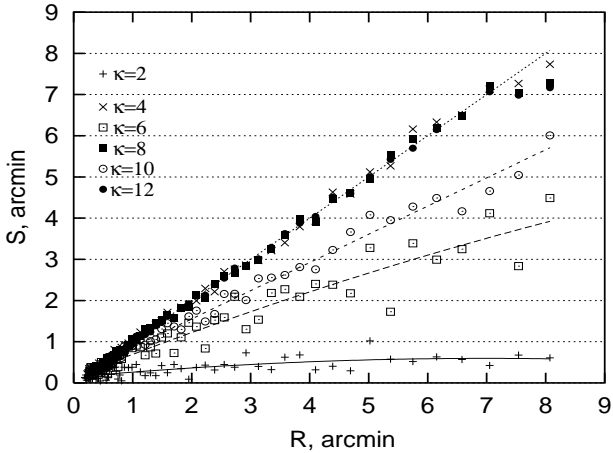}
\caption{  Effective field size $S_{\rm{rnd}}$ versus
radius $R$ of random samples at high star density; 
$k$ is the reference
group symmetry order; lines are data fits (\ref{eq:6_10})}
\end{figure}

To test the validity  of Eq.(\ref{eq:6_9}), we performed
numerical simulations that assumed the Galaxy model  
by Bahcall \& Soneira (1980). Random star samples
taken in the near-equatorial zone ($b=20{\degr}$, $l=0{\degr}$) to
V=23 mag were symmetrized, with condition (\ref{eq:4_10}), to
$k=2,4..12$ order and estimates $S_{\rm{rnd}}$ of $S$ for each
random sample were computed using Eq.(\ref{eq:4_12}). 
Random sizes $S_{\rm{rnd}}$ presented in Fig.9 are fitted with the function
\begin{equation}
\label{eq:6_10}
S_{\rm{exp}}=  	\chi R^{\mu}
\end{equation}
that gives the expectation of effective field size for 
a random star sample taken in a circle of a radius $R$; both 
$S_{\rm{exp}}$ and $R$ are given in arcminutes. 
The model (\ref{eq:6_10}) is consistent with (\ref{eq:6_9}) 
and extends this dependency to odd $k/2$. The 
quantities  $\chi$ and $\mu$  valid for both high and low
star density are given in Table 4. At polar
regions, the values of these parameters are higher, thus
slightly larger $S$ and $\Delta$ are expected at equal $R$.
\begin{table}[tbh]  %table 4         @{\extracolsep{1mm}}
\caption[]{ Coefficients of Eq.(\ref{eq:6_10}) for different star density}
\begin{flushleft}
\begin{tabular}{@{ }c|cc|cc} 
\hline
& \multicolumn{2}{c|}{ near-equatorial zone} & 
   \multicolumn{2}{c}{ Galactic pole}\\
\hline
  $k$  &  $\chi$  & $\mu$ &   $\chi$  & $\mu$ \\
\hline
 2  &  0.275 & 0.414 & 0.848 & 0.537   \\
 6  & 0.712 & 0.817 & 1.000 & 0.830  \\
 10 & 0.857 & 0.893 & 0.997 & 0.922 \\
\hline                                                    
%\end{tabular}
%\begin{tabular}{c}
 \multicolumn{5}{c} { $\chi= \mu =1$ for $k=4,8,12$ ($k/2$ even)} \\
\hline                                                    
\end{tabular}
\end{flushleft}
\end{table}

For $k/2$ odd the value of $S$ is a function of odd-order coordinate 
cross-moments averaged over the field. Result of averaging depends 
largely on
the contribution from a few bright stars with large $a_i$ and so 
is not returned to zero as it is expected for star fields with 
constant star brightness. Not performing a detailed study,
we assumed that in the spectral domain this effect is described by 
the empiric model 
\begin{equation}
\label{eq:6_11}
\begin{array}{ll}
Q(q) = & Q_{k+2}(q)+ (S/R)^k [J^2_{k/2}(2 \pi Rq) +\\
& +2 \sum \limits_{m=k/2+1}^{\infty}
J^2_{m}(2 \pi Rq)],  \; \: k/2 \; \: {\mbox{odd}}
\end{array}
\end{equation}
with correct asymptotic properties. 
Here $Q_{k+2}(q)$ is a  $k+2$ order function  
(\ref{eq:6_7});  $S$ is given either by Eq.(\ref{eq:6_10}) to produce
the mathematical expectation of the filter or is a sampled field size leading
to a "sampled" filter. The second
component in the expression is caused by incomplete averaging of odd-order
coordinate moments and decreases with $R$ since $(S/R)<1$. 
For fields with stars of equal brightness, this component vanishes
yielding  $Q(q) =  Q_{k+2}(q)$ and thus increasing the field 
symmetry order by +2.
It is easy to find that 
Eq.(\ref{eq:6_11}) at short $q$ follows approximation (\ref{eq:4_13}),
critically important for computation of $\Delta$. Direct calculations
of $\Delta$ for random star fields in Sect.9 had proved the validity
of the model  (\ref{eq:6_11}). 

Finally let us derive laws for  $\Delta$ 
as a function of   reference star number $N$ in the field,
and its radius.
As follows from Eq.(\ref{eq:4_9}), the field symmetry order
not can exceed $k_{\rm{max}}=\sqrt{8N+1}-1$.
Then, with some optional $k \leq k_{\rm{max}}$ and optimal apodization
 parameter
$\nu=k$ or $\nu=k+1$ (Table 1), we find that for a turbulent layer
at a height $h$
\begin{equation}
\label{eq:6_12}
\Delta^2 \sim \chi^{k} \left [ 
{\frac{hR}{3400D}}
\right ]        ^k R^{k(\mu -1)} D^{2/3}
\end{equation}
where $R$ is in arcminutes, $D$ and $h$ in meters. Consider that
at high star density, a $R=1'$ field is expected to contain about 
100 stars to
V=23 mag; using all of them as reference makes $k$ orders to 
$k_{\rm{max}} \approx 25$--30 quite feasible. A gain in $\Delta$
for narrow $hR <3400D$ fields is therefore huge.

%% file: 1481_6.tex
\section{Plate reduction}
An approach to plate reduction based on the current technique 
should take into account the fact that effective filtration of
atmospheric noise  occurs only in a two-component 
vector $\vec {W}$. As opposed to the common 
technique based on plate constant determination that 
provides
a global fitting of measured to model data at any point on the 
plate, the present
technique gives a local solution valid directly
at the point $x_0$, $y_0$. When multiple targets are considered,
each one should be processed with its own set of weights $a_i$
centered on the particular target. Below we explain 
in detail  the feasibility
of recovering positional information from  the  quantity $\vec {W}$,
taking into account the two possible goals of observations:

-- determination of the target object
position with reference to field stars, using their precise positions 
(a classic  problem of astrometry), and 

-- determination
of  the target proper motion with reference to field stars (positions
unknown) from observations  made in the two epochs. 

\subsection{ Compensation of low-order reduction model terms in $\vec {W}$ }
Consider polynomial expansion 
\begin{equation}
\label{eq:r_1}
\begin{array}{ll}
\bar{x}_i & = x_i+A_0+A_1^{1,0}(x_i-x_0)+A_1^{0,1}(y_i-y_0)\\
  	&+A_2^{2,0}(x_i-x_0)^2  +A_2^{1,1}(x_i-x_0)(y_i-y_0)\\
	&+A_2^{0,2}(y_i-y_0)^2+ \ldots =x_i+A_0         \\
	&+\sum \limits_{m=1}^{k/2-1} \sum \limits_{ \begin{array}{c}
					^{\alpha,\beta =0,}
					_{\alpha+\beta = m } 
					\end{array} }   ^m
	A_m^{\alpha,\beta }(x_i-x_0)^{\alpha}(y_i-y_0)^{\beta }\\
	&+ {\mathcal R}_{k/2}(x_i-x_0,y_i-y_0)
\end{array}
\end{equation}
of star  measured  coordinates $\bar{x}_i$, $\bar{y}_i$
over coordinate cross-moments $m=1,2 \ldots k/2-1$ of
actual coordinates $x_i$, $y_i$ 
not distorted by the atmosphere; the equation for $y$ has a 
similar structure and all coordinates are considered to be 
scaled to the turbulent layer height $h$. 
Expansion (\ref{eq:r_1}) is essentially
a reduction model used for fitting measured to standard coordinates. 
The first few modes of the expansion with amplitudes $A_m^{\alpha,\beta }$ 
represent common geometric terms: 
zero-point (mode $m=0$) and scale ($m=1$), also, they can be associated
with the influence of the classic optical aberrations: tilt ($m=0$), 
defocus and astigmatism ($m=1$), etc.

Each expansion term $A_m^{\alpha,\beta }$ may include also
a stochastic component of image motion which, 
due to a similar structure of Eq.-s (\ref{eq:r_1}) and (\ref{eq:4_7a})
is associated with a corresponding modal term of the $Q(q)$ function.
The model is truncated by some $k/2-1$ order; the sum of high $m \ge k/2$
modes is denoted as ${\mathcal R}_{k/2}$. 

An expression for the $W_x$ quantity is found by
subtracting from (\ref{eq:r_1}) an identity $\bar{x}_0=A_0+x_0$ valid
at the point $x_i=x_0$, $y_i=y_0$ and  performing a summation 
with  $a_i$: 
\begin{equation}
\label{eq:r_2}
\begin{array}{ll}
W_x & =\frac{1}{N}\sum a_i (\bar{x}_0-\bar{x}_i) \\
     &= -\frac{1}{N}\sum a_i {\mathcal R}_{k/2}(x_i-x_0,y_i-y_0) 
\end{array}
\end{equation}
Here all $A_1$, $A_2 \ldots A_{k/2-1}$  low-order
components vanish  due to conditions (\ref{eq:4_7a}).
The use of weights $a_i$  based on the true differential position of stars 
$x_0 - x_i$, $y_0 -y_i$ therefore 
allows us to form a linear combination (\ref{eq:r_2}) of measured 
coordinates which is insensitive to low-order terms of expansion 
(\ref{eq:r_1}), irrespective of whether they are constant or
stochastic ones. It is very important that $W_x$ does not depend on any
changes of aberrations and temporal variations of image motion
of low orders. The mathematical expectation of
$W_x$ thus is zero and its  variance  depends on high
$k/2$, $k/2+1 \ldots$ modes uncompensated in the
summation.

\subsection{Determination of target position}
Formulation of this particular problem implies that the precise position 
$x_{obj}$, $y_{obj}$ of the target object is unknown 
and $x_0$, $y_0$ is a preliminary object's position 
used for computation of weights $a_i$.
The true target position $x_{obj}=x_0 + \delta x$ ($\delta x$ is 
a correction to  $x_0$)
is related, as it follows from 
(\ref{eq:r_1}),  to a measured position
$\bar{x}_{obj}= x_0+ \delta x +
A_0+A_1^{1,0}\delta x+A_1^{0,1}\delta y+A_2^{2,0}\delta x^2+\ldots$. 
Then Eq.(\ref{eq:r_2}) compiled for $\bar{x}_{obj}$ instead of 
$\bar{x}_0$ becomes 
\begin{equation}
\label{eq:r_3}
\begin{array}{ll}
 {W_x}= & N^{-1}\sum _i a_i (\bar{x}_{obj}-\bar{x}_i) = 
\delta x+ A_1^{1,0}\delta x+ A_1^{0,1}\delta y \\
     &	+ \ldots - N^{-1}\sum _i a_i{\mathcal R}_{k/2} 
\end{array}
\end{equation}
Eq.(\ref{eq:r_3}) with a similar 
expression for $y$ and conditions (\ref{eq:4_7a}) for $a_i$ form
 a system with unknowns $\delta x$,  $\delta y$. In the first
approximation, $\delta x = N^{-1}\sum _i a_i (\bar{x}_{obj}-\bar{x}_i)$
yielding
\begin{equation}
\label{eq:r_3d}
{x}_{obj}={x}_0 + \bar{x}_{obj}-N^{-1}\sum a_i \bar{x}_i+ 
	N^{-1}\sum a_i {\mathcal R}_{k/2} 
\end{equation}
For large  $\delta x$, the system is solved by iterations, assuming $A_1=0$; 
weights $a_i$ are  recomputed after 
each refinement of  $\delta x$,  $\delta y$ and with a following shift
of the point $x_0$, $y_0$ to a new position. Iterations  converge very fast 
as $A_1 \ll 1$.

The above procedure allows us to obtain
precise positions, for example, of extragalactic radio sources
in the system of some high-accurate (future space mission) 
reference catalogue. 
Position of reference stars are used both for
determination of  $x$, $y$ frame origin (reference group zero point)
and for computation of $a_i$. Errors in reference star
positions affect computed $a_i$ values thus implicitly causing
a r.m.s. noise $\Delta_a$ in ${x}_{obj}$ positions (\ref{eq:r_3d}).
It is very important to know how large the  $\Delta_a$ component can be.

To study this effect, assume that  $\delta x_i$,  
$\delta y_i$  are random
and uncorrelated coordinate errors of the $i$-th star. Then the
use of $x_i+\delta x_i$,  $y_i+\delta y_i$ data 
yields biased estimates $a_i +\delta a_i$, 
which results in inaccurate compensation of atmospheric error
due to violation of conditions (\ref{eq:4_7}).
In particular, the first moment 
$M_x=\sum (a_i+\delta a_i)(x_i-x_0)$ 
is now not zero. Its value is easily found since
the second equation in (\ref{eq:4_7}) used for 
computations of weights now takes the form
$\sum (a_i+\delta a_i)(x_i-x_0+\delta x_i) = 0$.
Here
the second-order $\delta a_i\delta x_i$ terms can be discarded
yielding 
$M_x=-\sum (a_i+\delta a_i)\delta x_i \approx -\sum a_i\delta x_i $.
Therefore, the $q^2$ 
term of the $Q(q)$ function equal to  
$ 2(M_x^2+M_y^2)= 
2[(\sum a_i \delta x_i)^2+(\sum a_i \delta y_i)^2]$ 
in the left part of Eq.(\ref{eq:M_x}) becomes not zero. 
With respect to image motion statistics,
its influence is equal to 
that caused by a  $q^2$ term of the $Q(q)$ function for a stellar group
with effective coordinates $x'_i =a_i \delta x_i$, $y'_i =a_i \delta y_i$ 
and unit weights. 
This group, in turn, can be substituted by a double star having
a $Q(q)$ function  (\ref{eq:2_12}) with distance parameter 
$s_1=\sqrt{2} \sum a_i {\sigma}_i$ where 
${\sigma _i}$ is the mean coordinate error
of $i$-th star defined by equation 
$ \delta x_i^2+ \delta y_i^2=2 {\sigma}_i^2$.

Thus, inaccuracy in reference star coordinates results in the occurrence of 
an extra stochastic 
component whose behaviour is similar to the differential image motion of 
an "equivalent" double star with a very small separation $s_1$.
Assuming equal star brightness, 
and taking into account that $\bar{a^2} \approx (k/4)^2$ (Sect.6), 
we derive $ s_1^2 \approx k^2 N \bar{\sigma}^2/8$ where $\bar{\sigma}^2$
is the mean value of ${\sigma}_i^2$.

Consider, for instance, observations performed 
with a $D=100$ m telescope, $T=10$ min, $k=8$ and $\nu=9$ in 
$R=1'$  field located close to the galactic plane 
($b=20{\degr}$, $l=0{\degr}$), and with GAIA space catalogue
positions as reference. In this sky area, the expected number of stars
to the GAIA's limit V=20 mag is $N \approx 35$. Assuming $\bar{\sigma}
\approx 20$ $\mu$as as the average (Perryman et all. 2001)
leads to $s_1=0.34$ mas. Now, scaling 
Table 6 data with a dependency
$\Delta^2 \sim S^2$ valid for $k=2$ 
(\ref{eq:5_3}), we find that for a double star
system with $S=s_1 \sqrt{2}=0.48$ mas, the noise 
$\Delta_a$ is only $6 \cdot 10^{-4}$~$\mu$as. 
It is easy to ascertain that for any other parameters of observations
 $\Delta_a$ is always much smaller than the zero point error 
$\bar{\sigma}/ \sqrt{N}$.

\subsection{Determination of proper motions}
The most important applications for astrometry with large 
telescopes, of course,
are related to proper motion (exoplanet search) works. In this case
the plate reduction technique is aimed at  measurements of small
displacements $\Delta \mu$ of scientific objects which occur in the 
time interval between the two epochs $T_1$
and $T_2$. A frame of reference is given 
by the measured positions of field stars $\bar{x}_i$, 
$\bar{y}_i$. 

Extracting of high-accurate proper motion data is ensured by the next
procedure.

Observations of epoch $T_1$ are used for computation
of weights  $\bar{a}_i(T_1)$ which correspond to some $k$ 
symmetry  order. 
A virtual center of the frame is fixed at the observed 
target position $\bar{x}_0(T_1)$, $\bar{y}_0(T_1)$ of the object location 
(sometimes, to avoid confusion, the epoch of a quantity measured
is given in parentheses). 
Since true positions $x_i$, $y_i$ 
of stars are unavailable, computations are performed with the
modified system
\begin{equation}
\label{eq:B_2}
\begin{array}{lr}
\sum _i \bar{a}_i =N,       & \\
\sum _i \bar{a}_i (\bar{x}_i-\bar{x}_0)^{\alpha} (\bar{y}_i-\bar{y}_0)^
		{\beta}=0,& \; 
		\alpha + \beta = 1 \ldots  \frac{k}{2}-1
\end{array}
\end{equation}
using star positions shifted by the image motion. Solutions 
$\bar{a}_i(T_1)$  therefore differ from unbiased solutions 
$a_i(T_1)$ of  system (\ref{eq:4_7a}).
Then, with $\bar{a}_i(T_1)$ and  measured star coordinates 
$\bar{x}_0(T_2)$, $\bar{x}_i(T_2)$ at the epoch $T_2$, we form 
\begin{equation}
\label{eq:B_3}
\bar{W}_x= N^{-1}\sum _i \bar{a}_i(T_1)[\bar{x}_0(T_2) - \bar{x}_i(T_2)] 
\end{equation}
and a similar  $\bar{W}_y$ component of the vector 
$\vec{\bar{W}}$.  
Note that $\bar{W}_x$ is not zero as could be expected 
from (\ref{eq:B_2}) since  positions and $\bar{a}_i$
in Eq.(\ref{eq:B_3}) refer to different time moments.  

Definition (\ref{eq:B_3}) for $\bar{W}_x$ looks like 
Eq.(\ref{eq:4_1}) except for the use of $\bar{a}_i$ instead of optimal 
weights ${a}_i$ at which $M_x=0$ and $M_y=0$. Therefore a function
$Q(q)$ corresponding to $\vec{\bar{W}}$ contains
a small additional $q^2$ term which emerges since the first 
coordinate moments
$M_x=\sum \bar{a}_i(T_1)[x_0(T_2) - x_i(T_2)] $ and 
$M_y=\sum \bar{a}_i(T_1)[y_0(T_2) - y_i(T_2)] $
formed with the biased $\bar{a}_i(T_1)$ values are not zero.

To estimate $M_x$ and $M_y$ magnitudes, 
assume temporarily that $T_2$ refers to another time moment  of
the first epoch so that $\Delta \mu=0$, and 
consider a model
inversed to (\ref{eq:r_1}) and of a similar structure:
${x}_i  = \bar{x}_i+\hat{A}_0+\hat{A}_1^{1,0}(\bar{x}_i-\bar{x}_0)+
\hat{A}_1^{0,1}(\bar{y}_i-\bar{y}_0)+
	 \ldots + \hat{{\mathcal R}}_{k/2}$.
Assuming the model to be written for 
the moment $T_1$, subtracting  an expression 
$x_0=\bar{x}_0 + \hat{A}_0$ valid for the target object, 
performing summation with weights  $\bar{a}_i(T_1)$
and taking into account conditions  (\ref{eq:B_2}), we find
 $M_x=\sum \bar{a}_i (T_1)[x_i(T_1)-x_0(T_1)]=
\sum \bar{a}_i(T_1) \hat{{\mathcal R}}_{k/2}(T_1)$.
A similar expression, of course, is valid for the moment $T_2$
with $\hat{{\mathcal R}}_{k/2}$  related to $T_2$. The quantity $M_x$
thus is a stochastic
variable whose instantaneous value depends on a particular set of 
$\bar{a}_i$.
The mathematical expectation of  $M_x$ as zero is reached at $\bar{a}_i=a_i$, 
its variance  depends on the variance of  
the $ \hat{{\mathcal R}}_{k/2}$ term and
is therefore equal to the $x$-component of $\Delta^2$  
expected at some current $k$ and $\nu$. The average values
of   $M_x^2+ M_y^2$ are thus of the order of $\Delta^2$.

It follows that uncompensated extra image motion caused by a small $q^2$
component of $Q(q)$ can be approximated (see a similar discussion 
in Section 7.2) by the image motion
in the "equivalent" double star system with a separation
$s_1 =\sqrt{M_x^2+M_y^2}=\Delta^2$.
With $\Delta$ given in Table 6 one  can
evaluate $s_1$ for specific parameters of observations; 
thus for a 10 m telescope  we find $s_1 \le 0.1$ mas if $k \geq 4$.
Image motion induced in this double star system is very
weak and normally can be disregarded.

The use of measured star positions for computation
of weights thus maintains
the high accuracy of the method.

In the above analysis  $T_2$ was related to the first epoch 
to null a proper motion effect. Putting  $T_2$ in the second epoch
presents some problems since now 
$x_0(T_2)=x_0(T_1)+ \Delta \mu _x$. Using a reversed model of 
 measured to standard coordinates transform, we obtain
the relation $x_0(T_2) =\bar{x}_0(T_1)+\bar{ \Delta \mu _x}
+\hat{A}_0+\hat{A}_1^{1,0}\bar{ \Delta \mu _x}+
\hat{A}_1^{0,1}\bar{ \Delta \mu _y}+ \ldots$  where
$\bar{ \Delta \mu _x}$ and $\bar{ \Delta \mu _y}$ are the measured
$x$ and $y$ components of  $\Delta \mu $. 
Assuming for field stars $x_i(T_2)=x_i(T_1)$, 
taking into account above expression
and Eq.(\ref{eq:B_2}), we find
\begin{equation}
\label{eq:B_W}
\begin{array}{l}
N^{-1}\sum \bar{a}_i(T_1)[\bar{x}_0(T_2) - \bar{x}_i(T_2)] \\
        =\bar{ \Delta \mu _x}+ \hat{A}_1^{1,0}\bar{ \Delta \mu _x}
      +\hat{A}_1^{0,1}\bar{ \Delta \mu _y}+ \ldots 
	 +N^{-1}\sum \hat{{\mathcal R}}_{k/2}
\end{array}
\end{equation}
This equation looks like Eq.(\ref{eq:r_3}) but cannot be solved
by the previously described iterations which involve refinement of weights 
since $\bar{a}_i$ 
are centered at a fixed point $\bar{x}_0(T_1)$, $\bar{y}_0(T_1)$. 
However, a simple truncation of second-order terms 
$\hat{A}_1^{1,0}\bar{ \Delta \mu _x}$
and $\hat{A}_1^{0,1}\bar{ \Delta \mu _y}$
 and substitution of the measured 
$\bar{ \Delta \mu _x}$ for the true  $ \Delta \mu _x$ displacement  yields
\begin{equation}
\label{eq:B_M}
\Delta \mu _x =N^{-1}\sum \bar{a}_i(T_1)[\bar{x}_0(T_2) - \bar{x}_i(T_2)] 
-N^{-1}\sum \hat{{\mathcal R}}_{k/2}
\end{equation}
The error caused by neglecting the difference between 
$\bar{ \Delta \mu _x} $ and $ \Delta \mu _x$ is small 
and has a variance equal to that  
of image motion in a double star system with 
$s_1= \Delta \mu _x$; it can be safely ignored
for small $\Delta \mu $. For 
instance, for distances $\Delta \mu=4$ mas  measured
with a $D=10$ m aperture, any $k$, $\nu=3$ and $T=10$ min
the bias is about $340''\cdot 10^{-6}\sqrt{\Delta \mu /40''}\sim 3.4$ 
$\mu$as. The estimate was found by scaling the value of 
$\Delta$ given in Table 6 for  a double star, from $s_1=40''$ 
to the length $ \Delta \mu $ with a dependency 
$\Delta^2 \sim S^2$ .
Even for such a large displacement, the relative error is about 
$10^{-2}$.

The reduction technique considered in this Section thus  
ensures extracting of proper motions to within $\Delta$ accuracy,
assuming of course that there are no  other sources of noise. 

%% file: 1481_7.tex
\section{Image motion integrated over the atmosphere}
\subsection{The model of $C_n^2$ vertical profile}
Because a total variance $\Delta^2$ of image motion is equal to a sum
of $\Delta^2(h)$ additives generated by each turbulent layer, its value
therefore is a function of the $C_n^2(h)$ vertical profile.
In Fig.10 we reproduce typical
plots of averaged $C_n^2(h)$ for the three Chilean sites: Cerro
Tololo ({\it http://www.gemini.edu/sciops/instruments/
adaptiveOptics/Seeing.html}), 
Cerro Paranal (average over all profiles in Fig.2 
given by Loaurn et al. 2000) and San Pedro Martir (Avila \& Vernin 
1998; average for the 1.5 and 2.1 m telescopes). 
The model profile by Hufnagel (1970) is shown for comparison.
Since the data for $h>20$ km are often unavailable
in original papers, we extrapolated $C_n^2$ up to 
$h=25$ km using a scaled profile from Serro Paranal (measured to
25 km height) matched to
the measured data  at $h=20$ km for the other sites. 

The divergence between local averaged profiles in Fig.10  
is about a half of a decade. However,
the temporal variations of $C_n^2(h)$ local profiles are much stronger,
which Loaurn et al. (2000) had demonstrated in Fig.2 for the
Cerro Paranal site. 
For these reasons it seems impossible to suggest 
a universal model of $C_n^2(h)$ that will adequately match
the real shape of the turbulence profile for any atmospheric conditions 
even at a single place. 

To derive numerical estimates, we defined the model of $C_n^2(h)$
as an average of San Pedro and Cerro Tololo data which represent
high and low limits of $C_n^2$ at $h>20$ km. 
Of course, the adopted model of $C_n^2(h)$ is somewhat arbitrary
and, due to varying atmospheric conditions, may give a 
factor 3--5 incorrect predictions for a sample value of $\Delta$. 
The model nevertheless gives quite reliable estimates of average
$\Delta$ for the Chilean sites and thus
sufficiently well serves for the purpose of this discussion.
The vertical profile of the wind velocity $V(h)$ (Table 5) 
which represents mean conditions for the South Geminy Telescope
(Cerro Pachon, Chile) was taken from Avila et al. (2001).
\begin{table}[tbh]  %table 5
\caption[]{ Wind velocity model taken from  Avila et al. (2001)}
\begin{flushleft}
\begin{tabular}{@{ }cc@{ }} 
\hline
  $h$, km  &  $V$, m/s \\
\hline
   $<7$  &  10   \\
   7--12 & 20 \\
   12--17 & 40 \\
   17--19 & 20 \\
   $>19$ & 10    \\
\hline                                                    
\end{tabular}
\end{flushleft}
\end{table}

\begin{figure}[tbh]  %fig. 10
\includegraphics*[ width=8.3cm, height=6.2cm]{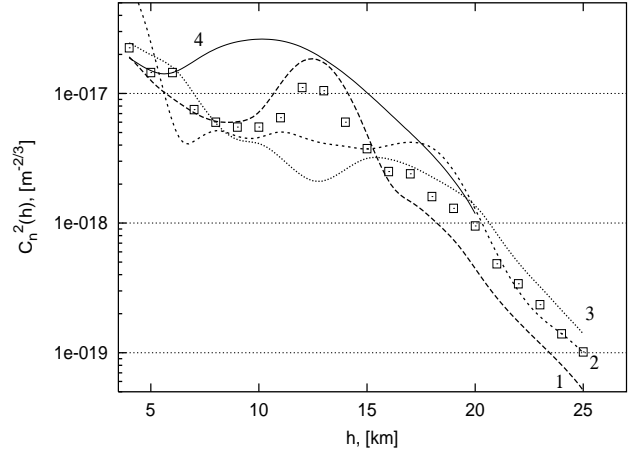}
\caption{  Average $C_n^2(h)$ profiles for Cerro Tololo (1),
Cerro Paranal (2), San Pedro Martir (3), Hufnagel's model (4) and
   the present model (squares)}
\end{figure}

\subsection{Contribution from different altitudes}
For narrow fields $\rho h \ll D/2$, application of the current method
of differential observations is very promising since $\Delta$
goes now as a power 
$ k/2$ (if $k<\nu +p$) or
$ (\nu +p)/2$  (if $k>\nu +p$) of the small quantity $\rho h /D$.
For a 100 m telescope, the  narrow field condition holds 
at any $h \le 30$ km  providing the effective size $\rho$  does not exceed 
$1-2'$. For this reason, any increase of $k$ and $\nu$ parameters,
corresponding to  movement down the Table 1 diagonal, 
always results in a better
suppression of turbulent effects, especially those generated at low altitudes. 
In the case of a 10 m telescope, the choice of $\rho$ is critical
for the validity of the narrow field condition as even 
at quite moderate $\rho=1'$ it turns to be 
violated already at about $h =15$ km. For upper layers, 
a much flatter, less efficient power dependency with 
an index $p/2$ holds, signalling a turn to
a wide field mode of differential measurements (Lazorenko 2002a).

\begin{figure}[tbh]  %fig. 11
\includegraphics*[ width=8.3cm, height=6.2cm]{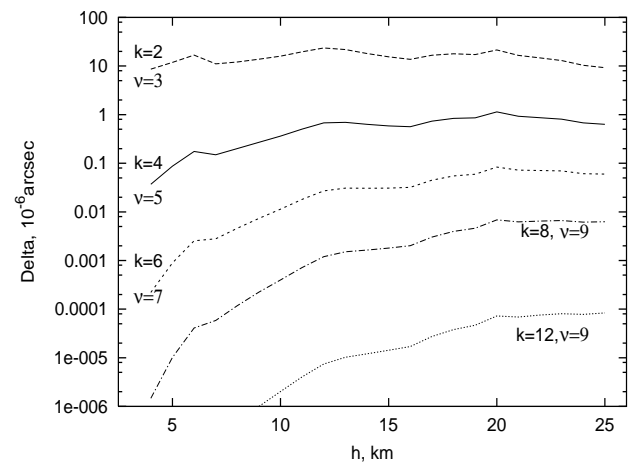}
\caption{ A contribution  $\Delta(h)$ from each 1 km thick 
atmospheric layer as a function of $h$, $k$ and $\nu$ 
for a 100 m telescope,
10 min exposure and $1'$ effective field radius}
\end{figure}
\begin{figure}[tbh]  %fig. 12
\includegraphics*[ width=8.3cm, height=6.2cm]{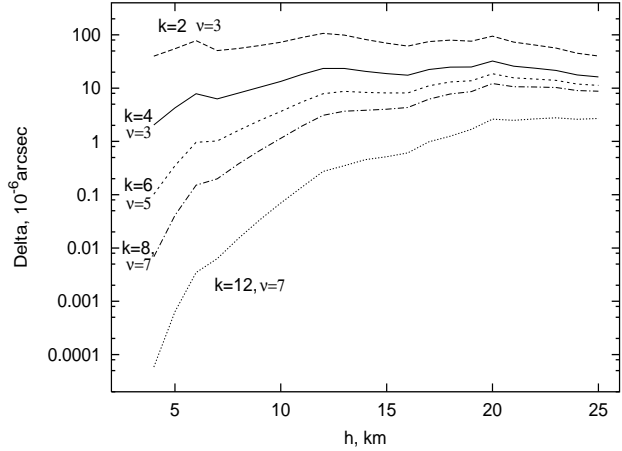}
\caption{The same as in Fig.11 for a 10 m telescope   }
\end{figure}

The model  of  $C_n^2(h)$ defined in the above subsection was used 
to compute the contribution $\Delta(h)$ from each $ \Delta h =1$ km 
turbulent layer  to the total value of $\Delta^2$. 
The plots in Fig.-s 11 and 12 display the function $\Delta (h)$ 
computed for  a 100 and 10 m telescopes 
placed at $h_0=2.5$ km altitude, 
$T=10$ min,
$p=2/3$ and  stellar reference groups "a, b, d, g, f" of Table 2, with 
 $k$ ranging from 2 to 12. The apodization parameter
was near optimal  $\nu=k+1$. Compare the effect of high order
symmetrization to the simple $k=2$ (Fig.11);
for instance, with $k=8$ and altitudes $h<10$ km the value of 
$\Delta (h)$ is  5 orders lower than that with $k=2$. 
At $h \sim 20$ km the gain still is 3 decades
if $D=100$ m. 

While a contribution  
from low-altitude layers is large
for asymmetric groups ($k=2$), symmetric groups  with  
$k>4$ orders are less sensitive to the turbulence at $h<15$~km.
The integrated value of $\Delta$ for high $k$  depends largely
on high altitude turbulence in spite of the fast decrease of  $C_n^2$
with $h$. 
For $k>8$, the function $\Delta(h)$  increases
until $h=25$ km, at least for the  $C_n^2$ profile adopted. The
data on  $C_n^2$ behaviour 
at $h \sim$ 25--40 km therefore is of special interest
for the correct prediction of the image motion variance. Probably,
the estimates of $\Delta$ computed in this study for 
$k \ge 8$ are slightly underestimated due to absent  data for $h>25$ km.

\subsection{The integrated variance of image motion}
Table 6, similar to Table 1, gives values of
$\Delta$ integrated over the atmosphere (in zenith direction)
for apertures $D=4$, 10, 30 and 100 m. Calculations have been
performed with the $C_n^2(h)$ and $V(h)$ model described above,
$p=2/3$, $T=10$ min and telescope altitude $h_0=2.6$ km. 
The $k$ values ranging from
2 to 12 refer to tutorial configurations of Table 2, except 
the example of $k=4$ order at  entry "c.f." that presents a non-discrete 
stellar field with the $Q(q)$ function (\ref{eq:6_7}). 
Since the effective
angular size $\rho=1'$ of each group corresponds to $S \approx 4$
-- 6 m at  15 -- 20 km height, the condition of very narrow
field $S \ll D/2$  is met only for $D=100$ m, and,
partially, for a $D=30$ m. 

\begin{table}[tbh]  %table 6
\caption[]{The variance $\Delta$ ($\mu$as)
integrated over the atmosphere
for  star configurations "a--h" of Table 1 and continuous field "c.f."
 for some $k$ and $\nu$ parameters; $T=10$ min;  $\rho=1'$} 
\begin{flushleft}
\begin{tabular}{@{}c@{}|rrrrrrrr@{}}
\hline
      &  \multicolumn{8}{c}{ $ k$ }     \\
\cline{2-9}
   & 2  &  4   & 4     & 4 & 4    &  6    &  8    & 12  \\
\cline{2-9}
      &  \multicolumn{8}{c}{  conf. name}      \\
\cline{2-9}
   &  a &   b    &  c  & f  & c.f. &  d     &  g     &  h   \\
\hline
$\nu$      &  \multicolumn{8}{c}{ $ D=4$ m}     \\
\hline
 3 &  572 &   261 &244&   254 & 233 &   167 &   114 &   44             \\  
 5 &  634 &   315 &293&   305 & 276 &   219 &   159 &   58             \\  
 7 &  679 &   356 &330&   344 & 307 &   263 &   200 &   83              \\ 
 9 &  710 &   387 &359&   374 & 332 &   297 &   235 &  106             \\  
\hline     
$\nu$    &  \multicolumn{8}{c}{ $ D=10$ m}     \\
\hline
 3 &  337&    89& 87 &    88&   85&   40 &  25    & 10.6          \\  
 5 &  382&   114&110 &   112&  109&  46 &  21    & 4.9           \\  
 7 &  419&   139&134 &   137&  132&  63 &  29    & 4.6            \\ 
 9 &  448&   162&155 &   159&  151&  80 &  41    & 7.0           \\  
\hline     
$\nu$      &  \multicolumn{8}{c}{ $ D=30$ m}     \\
\hline
 3 &   165&  19  & 19 &  19 & 19 &   7.4&  4.80  & 2.03          \\  
 5 &   188&  21  & 22 &  21 & 22 &   3.9&  1.27  & 0.26          \\  
 7 &   207&  27  & 27 &  27 & 27 &   4.8&  0.98  & 0.08           \\ 
 9 &   224&  32  & 33 &  32 & 33 &   6.4&  1.30  & 0.05          \\  
\hline     
$\nu$     &  \multicolumn{8}{c}{ $ D=100$ m}     \\
\hline
 3 &      74&   3.5& 3.5 &  3.5&3.4&  1.21&   0.783&   0.3038    \\  
 5 &      84&   2.9& 2.9 &  2.9&3.0&  0.21&   0.061&   0.0111    \\  
 7 &      92&   3.6& 3.7 &  3.6&3.8&  0.20&   0.017&   0.0010     \\ 
 9 &      99&   4.5& 4.6 &  4.5&4.7&  0.27&   0.018&   0.0002    \\  
\hline                                                    
$N_{min}$ &  1 &    3   &  3 & 3 & $\infty$ &   6    &  10    &   21  \\
\hline                                                    
\end{tabular}                                          
\end{flushleft}                                        
\end{table}

The estimates given for a 100 m telescope confirm the
efficiency of the high $k$ and $\nu$ used. In comparison to normal
observations ($k=2$, $\nu=3$), the gain in $\Delta$ is
about 5 orders of magnitude with extreme parameter  values given in
Table 6. The improvement in $\Delta$ which occurs 
with an increase of $k$ at any $\nu$ is also typical. On the contrary, 
the increase
of $\nu$ at fixed $k$ is useful only up to $\nu \sim k$. A similar
dependency is valid for smaller apertures, with a tendency of 
the optimal $\nu$ value to decrease to $\nu=k-1$ or even $\nu=3$ 
(no apodization) for a 4 m telescope. 
Note the very fast increase of $\Delta$ at transitions to smaller 
apertures.
Nevertheless, considering entry for a 10 m telescope, one can note
a progressive improvement in $\Delta$ with implementation of high
$k$. Even with no apodization, the atmospheric error can be reduced
to 10 $\mu$as providing that at least 21 reference stars are
available in a circle of $1'$ radius (see Eq.(\ref{eq:4_9})). 
Even limited to 6
stars, which does not allow $k$ larger than 6, one can expect 
still quite small
($\sim 40$ $\mu$as) errors suitable for 
exoplanet search programs.

Symmetric "b, f", strongly asymmetric
"c" and non-discrete "c.f." configurations 
of equal $k=4$ order have been purposely included in 
Table 6 to show that the magnitude of
$\Delta$  depends rather weakly on peculiar features and type
of star distribution in  the field, and is, in fact, a function of
$k$ and $\nu$ providing that $\rho$ is fixed.

\begin{table}[tbh]  %table 7
\caption[]{ The variance  $\Delta$ ($\mu$as)
for the star field of Table 3 at 10 min exposure} 
\begin{flushleft}
\begin{tabular}{c|rrrr}
\hline
      &  \multicolumn{4}{c}{ $ k$ }     \\
\cline{2-5}
   & 2          &  4       &  6    &  8    \\
\hline
$\nu$    &  \multicolumn{4}{c}{ $ D=10$ m}     \\
\hline
 3 &   60      &  53      & 14      &   9.0       \\  
 5 &   75      &  67      & 13      &   5.8       \\  
 7 &   91      &  82      & 18      &   7.8        \\ 
 9 &  105      &  96      & 23      &  11.1       \\  
\hline
$\nu$    &  \multicolumn{4}{c}{ $ D=100$ m}     \\
\hline
 3  &    5.1    &   2.1    &   0.420  &   0.268     \\  
 5  &    5.5    &   1.7    &   0.050  &   0.016     \\  
 7  &    6.2    &   2.2    &   0.042  &   0.004      \\ 
 9  &    6.8    &   2.6    &   0.056  &   0.004     \\  
\hline                                                    
$\rho$ &   3.7$''$ & 45.7$''$ &   35.2$''$ &  42.2$''$ 	\rule{0pt}{11pt} \\
\hline                                                    
\end{tabular}                                          
\end{flushleft}                                        
\end{table}

Table 7 represents estimates of $\Delta$ for a reference
field given in Table 3, at 10 min exposure; the last line
contains effective sizes $\rho$ 
for each $k$. Note  that a good quasi-symmetric
distribution of stars  alleviates the difference in $\Delta$
between  $k=2$ and $k=4$ orders,  
the noise for $k=2$ is  only slightly over that given for $k=4$.

Pravdo \& Shaklan (1996) derived an estimated $\Delta =150$ $\mu$as/hr 
for the magnitude of atmospheric fluctuations at the 5 m telescope 
and Table 3 star field. This is a value 
obtained by considering each field star, in turn, 
as a target, and by averaging individual 
estimates of   $\Delta$ that fluctuated at least a factor of 2--3.
Following the target position
changes, the effective frame size (approximately equal to the frame 
radius  with the target at its center) varied from 3.7$''$ to
90$''$ for the outer stars. For this range of $\rho$, our model
predicts variations from  55 to 300 $\mu$as/hr which well matches 
the observed value of 150 $\mu$as/hr.

%% file: 1481_8.tex
\section{Astrometric performance of very large ground-based telescopes}
Realization of   1--10 $\mu$as accuracy
 requires a good elimination of various noise sources related to
optical aberrations,  pixelization effects (especially 
for small images produced by adaptive telescopes), 
photon noise in  star images, differential
chromatic refraction (DCR)  etc. As it was noted by Louarn et al. (2000),
in particular, the  problems caused
by a DCR  that stretches the star images  into colored
strips are very intricate. 
The amplitude of the DCR effect depends
on zenith distance, air temperature and pressure, spectral band
 and star colors.
Using Allen's (1973) tables, one can find that two rays with 
 wavelengths of 500 and 600 nm coming from a star 
at a zenith distance of $30{\degr}$ are imaged with a separation of about 
180 mas along  a vertical direction. 
The noise  induced by this effect in the differential position of stars
(Pravdo \& Shaklan 1996)
amounts to about 60 $\mu$as in a 1.5$'$ field  for the 5-m Palomar telescope. 
Fortunately, in proper
motion  studies the DCR effect is residual and essentially weakened 
since star motions  are found from residuals of differential 
star positions in the two epochs. 
The modelling  of DCR based on use of atmospheric bulk parameters
is  therefore very promising.
We have found that
a proper control of atmospheric air
parameters (air temperature to 0.2${\degr}$,
pressure to 0.2~mb) allows one to apply corrections which reduce 
DCR noise  to 8 $\mu$as in
the relative displacement  of  A and M stars in the field of
1.5$'$. Once effective wavelengthes of stars are known to 0.4 nm, 
the noise decreases to 0.8 $\mu$as.
It should be noted that for high quality adaptive optics producing images
with FWHM$\ll 100$ mas, the DCR corruption 
of images is so strong that it makes them entirely unsuitable for 
measurements.
It is necessary therefore to use some special optics  
for compensation of atmospheric chromatism, otherwise the filter width
should be strongly narrowed.

Assuming that solution of this and other problems will be eventually
found by progresses in technology, we restrict the error budget
with two components: the atmospheric image motion and photon
noise with variances $\Delta^2$ and $\sigma_{ph}^2$ respectively.
The contribution from both effects was evaluated as a function of
the angular field size $R$ for  sky
star densities near the galactic plane and at the pole.
All particular cases of aperture, image parameters, exposure,
field size etc. of course, not can be considered; therefore
we restricted analysis only to the case of a future extremely large
100 m telescope and modern 10 m class telescopes.

Estimates of $\sigma_{ph}$ were found with use of Eq.(\ref{eq:6_6b}).
The value of FWHM in this expression strongly depends
on the performance of adaptive optics, the  telescope
aperture and may vary from 0.0015$''$ (diffraction limit of a 
100 m telescope) to 0.4$''$ (atmospheric uncorrected seeing).
The next estimates for a  100 m telescope assume FWHM=0.1$''$ 
achievable with low-order adaptive optics, and for a 10 m telescope
a quite conservative FWHM=0.4$''$ was adopted.  We assumed 
that observations are obtained
in zenith, in R band, CCD quantum
efficiency  0.85, transmission of optics 0.8 and of atmosphere
0.9.   Then a star 
of V=15 mag and of average spectral type, being observed with a 100 m 
telescope, will provide $0.93\cdot 10^{8}$ 
detected electrons/sec (Allen 1973).
A total light $\sum n_i$ of the star field 
was estimated based on the Galaxy model by
Bahcall \& Soneira (1980). Stars fainter than V=23 mag were not
considered as they give low light signal. To obtain more robust results, 
the expected star number in each 1 mag bin was rounded (truncated) to
the smaller integer. This procedure trimmed the bright
end of stellar magnitudes due to which   very narrow star
fields were formed largely by the faintest stars.

\begin{figure}[tbh]  %fig. 13       height=8.4cm
\includegraphics*[ width=8.8cm, height=11.4cm]{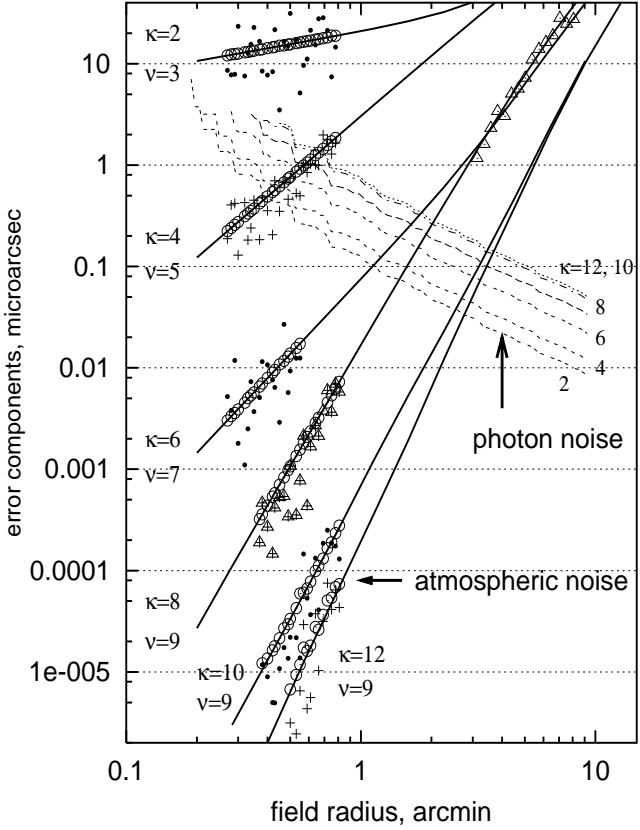}
\caption{Atmospheric and photon  noise as a function
of field radius $R$ for a 100 m telescope, 10 min exposure,
near the Galactic equatorial plane and a set of $k$ and $\nu$ parameters. 
Atmospheric noise: dots -- for
random star fields  ($\Delta_{\rm{rnd}}$) at $k/2$ odd; 
crosses -- the same at $k/2$ even;
solid lines -- expected mean noise  $\Delta_{\rm{exp}}$ for "typical" 
stellar fields;
circles -- transformation (\ref{eq:9_1}) from   $\Delta_{\rm{rnd}}$
to $\Delta_{\rm{exp}}$ (see text);
triangles -- reversed transformation.  Dashed lines -- 
photon noise $\sigma_{ph}$  at $0.1''$ seeing }
\end{figure}

 Fig.13 represents data 
for a 100 m telescope, $T=10$ min, galactic coordinates $b=20{\degr}$, 
$l=0{\degr}$ and $k$ in the range from 2 to 12.  For $k \le 8$,
the apodization parameter was set to be $\nu=k+1$, for higher $k$
its value was limited by $\nu=9$ so as not to worsten the light 
transmission. The dashed lines represent $\sigma_{ph}$ 
computed with Eq.(\ref{eq:6_6b}); plots start from the smallest field size
which ensures $N_{min}$ star number necessary to
realize $k$ order symmetry.

The estimates of $\Delta$ were computed by integration of 
Eq.(\ref{eq:2_11}) for the turbulence model described in 
Section 8. It was performed
two ways: 

\noindent  1) by processing random
stellar fields simulated with a Galaxy model (Bahcall \& Soneira 1980). 
It involved calculation  
of $a_i$ defined by conditions (\ref{eq:4_7}), (\ref{eq:4_10}) 
and  of $Q(q)$ filter function (\ref{eq:4_4}) for each  field. 
These direct point estimates $\Delta_{\rm{rnd}}$ of 
atmospheric error $\Delta$ computed for $R<0.8''$ are shown by 
dots for odd $k/2$ and by crosses for $k/2$ even. 

\noindent 2) by computing the atmospheric error for "typical" stellar fields
whose filter $Q(q)$ mathematical expectation is given by 
Eq.(\ref{eq:6_11})
for odd $k/2$ (parameters $\chi$ and $\mu$ taken from Table 4)  and  
(\ref{eq:6_7}) for even $k/2$. 
Computed estimates of $\Delta_{\rm{exp}}$ are shown by solid lines.

The difference between  $\Delta_{\rm{rnd}}$ and   $\Delta_{\rm{exp}}$ 
estimates is caused by  different values of sampled effective sizes 
$S_{\rm{rnd}}$ and of 
$S_{\rm{exp}}$ (the expected effective size).
The inter-relation of these values is given by the expression
\begin{equation}
\label{eq:9_1}
\Delta_{\rm{exp}}= \Delta_{\rm{rnd}}( S_{\rm{exp}}/S_{\rm{rnd}})^{k/2}
\end{equation}
that follows from  the power law (\ref{eq:5_3}) 
valid for $k<\nu+p$.
Open circles in Fig.13, the results of transformation (\ref{eq:9_1}) 
from $\Delta_{\rm{rnd}}$
to $\Delta_{\rm{exp}}$, are shown to be placed perfectly along solid 
lines that represent $\Delta_{\rm{exp}}$.

The expression opposite to Eq.(\ref{eq:9_1}) can be used for
indirect computation of $\Delta_{\rm{rnd}}$ proceeding from 
an effective frame size $S_{\rm{rnd}}$ and  $\Delta_{\rm{exp}}$ 
for fields containing
$N \sim 10^2$--$10^3$ stars. Direct 
numeric integration
in this case is too time-expensive since the number of terms
in Eq.(\ref{eq:4_4}) increases as $N^2$ or $R^4$. With this approach,
atmospheric noise is easily estimated at any large $R$.
Estimates of
$\Delta_{\rm{rnd}}$ computed based on $S_{\rm{rnd}}$ and 
$\Delta_{\rm{exp}}$ are  shown in Fig.13
by triangles for $k=8$, $\nu=9$ and $R$ varying from 3 to $8'$ when star
number in the field mounts from 1000 to 3000. For $R<0.8'$, approximate 
estimates
(triangles) exactly match those directly computed $\Delta_{\rm{rnd}}$ 
(crosses).

The positive factor 2 offset of $\Delta_{\rm{exp}}$  over point
estimates $\Delta_{\rm{rnd}}$ seen for $k=4,8,12$ and very narrow
$R \leq 0.6'$ fields with $N \leq 20$, originates from use of
a non-descrete field model for computation of $\Delta_{\rm{exp}}$,
or the assumption $S=R$ (\ref{eq:6_9}). 
At very low $N$, however, this gives rather
an upper limit of $S$ but its mathematical expectation since
stars do not entirely cover periphery of 
the field. The difference is small
but becomes apparent after being amplified  due to the power dependency
$\Delta \sim S^{k/2}$.

A short comment should be made concerning plots for $k=10$ and $k=12$
with the noticeable scattering of $\Delta_{\rm{exp}}$ (circles) 
computed with Eq.(\ref{eq:9_1}). 
For these plots, observations are carried out under condition 
$k>\nu+p$, which, according to  Eq.(\ref{eq:5_2}), means that
$\Delta^2$ is proportional to
the power $\nu+p$ of  modified frame size $ \hat{S}$ which
is not equal to $S$.
Though Eq.(\ref{eq:9_1}) is not valid here,  it still
gives rather good results.

\begin{figure}[tbh]  %fig. 14       height=8.4cm
\includegraphics*[ width=8.8cm, height=7.4cm]{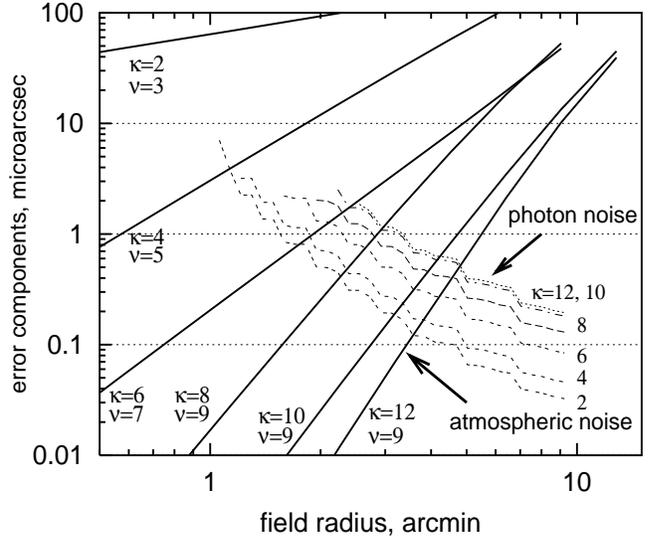}
\caption{ The same as in Fig.13 ($D=100$ m), for
the Galactic pole; data for random star fields are not shown}
\end{figure}

A plot with error estimates for the Galactic pole 
is given in Fig.14.  For a 10 m telescope, results are represented
in Fig.-s 15, 16; no apodization is applied since it does not
offer an improvement (Table 6).

\begin{figure}[tbh]  %fig. 15       height=8.4cm
\includegraphics*[ width=8.8cm, height=7.4cm]{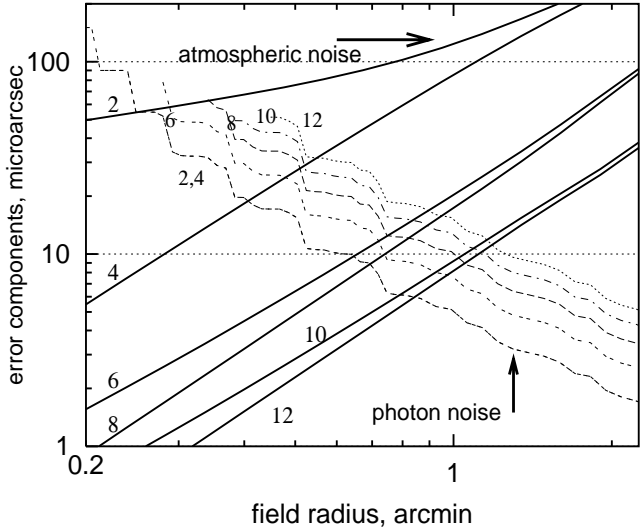}
\caption{ The same as in Fig.13 (Galactic equator), for
a 10 m telescope and $0.4''$ seeing; data for random star fields 
are not shown. No apodization ($\nu=3$) }
\end{figure}

\begin{figure}[tbh]  %fig. 16       height=8.4cm
\includegraphics*[ width=8.8cm, height=7.4cm]{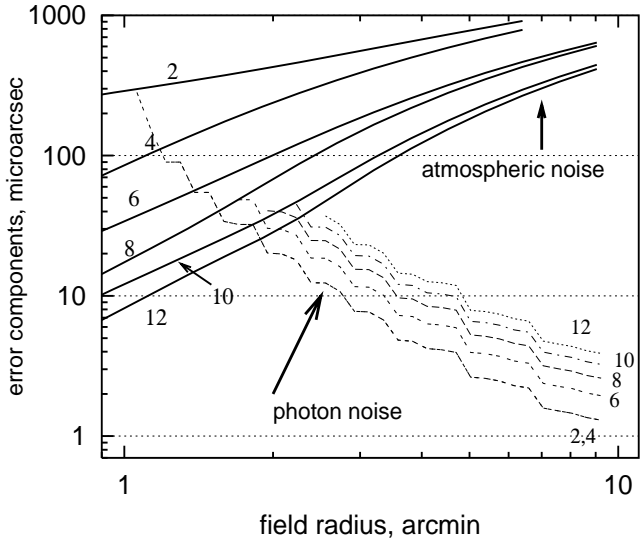}
\caption{ The same as in Fig.15 (a 10 m telescope) for a Galactic pole} 
\end{figure}

A peculiar feature of Fig.-s 13--16 is the critical point
$R_0$ at which plots of $\sigma_{ph}$ and $\Delta$ drawn for a certain
$k$ and $\nu$ intersect.
For $R<R_0$, where filtration of atmospheric noise is very efficient,
photon noise dominates over
$\Delta$ (photon-limited observations). For $R>R_0$, on the contrary,
 $\Delta > \sigma_{ph}$ and observations are atmosphere limited.
At low sky star density, reference frames
containing at least $N_{min}$ stars (brighter than V=23), necessary
to form  $k$ order symmetry fields, can  be formed
only at relatively wide  $R$ where $\Delta$ becomes large. The typical
situation described is shown in Fig.16 (polar regions) where plots for 
$\sigma_{ph}$ and $\Delta$ expected for a 10 m telescope
do not intersect at any $R$, signifying that
observations are atmospherically limited.

\begin{figure}[tbh]  %fig. 17       height=8.4cm
\includegraphics*[ width=8.8cm, height=7.4cm]{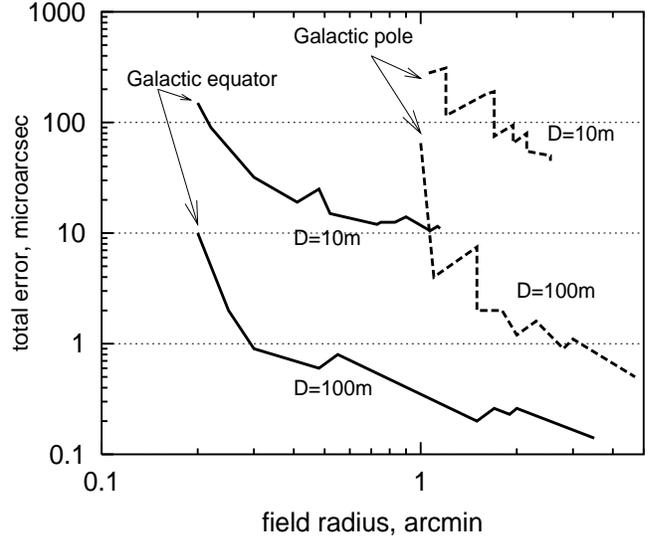}
\caption{ Total optimized atmospheric and photon noise error $\sigma_{t}$
as a function of field radius $R$ for 10 and 100 m telescopes,
a 10 min exposure and parameters
of observations same as in Fig.-s 13--16; solid lines --- for Galactic equator
($b=20{\degr}$, $l=0{\degr}$); dashed -- for the Galactic pole. 
Along each line, $k$  value increases with $R$ from 2 to 12} 
\end{figure}

Consider now a total error $\sigma_{t}$ which we define  simply as 
being equal to the largest error component  $\sigma_{ph}$ or $\Delta$.
At any $R$, a value of  $\sigma_{t}$
can be minimized by proper selection of $k$ and $\nu$.
The plot of optimal $\sigma_{t}$ error 
as a function of $R$ was formed taking the best combination of  segments
of $\sigma_{ph}$ and $\Delta$ curves in 
Fig.-s 13--15. Resulting segmented curves drawn for 
the cases discussed are  shown in Fig.17. 
Each curve starts at small $R$ where only low order $k=2$ or $k=4$ 
symmetry is realized
with 1--3 reference stars; the right ends of curves correspond to $k=12$
and fields containing $N \sim 100$--1000 stars (except for the case of
a $D=10$ m telescope operating in the polar region where $N \approx N_{lim}$).
Plots show that $\sigma_{t}$ is a decreasing function of $R$, so use of high
$k>12$ orders results in a progressive slow decrease of $\sigma_{t}$,
which, however, involves the use of a wider field. From Fig.17 data we
may notice that the
near optimal field size is 0.4--1$'$ at high star density and about
2$'$ at polar regions.

With above assumptions on FWHM, and at optimal field size, the expected
error of ground-based observations for a 10 m telescope (no apodization) 
is, depending on sky star density,
10 to 60 $\mu$as per 10 min exposure. For a 100 m telescope this estimate
is 0.2 to 2 $\mu$as.

%% file: 1481_9.tex
\section{Conclusion}
As commented by Lindegren (1980), dependency 
$\Delta \sim (\rho h /D)D^{1/3}T^{-1/2}$ valid for conventional
differential astrometric observations in narrow fields
(in wide fields $\Delta \sim (\rho h )^{1/3}T^{-1/2}$)
is not a fundamental limitation to observations from the ground
as it can be improved by applying a new method of measurements.
In this paper we introduce a method that essentially impairs 
restristions on the precision of ground-based astrometry 
caused by atmospheric turbulence; the power dependency of 
$\Delta$ on $\rho h $, $D$ and $T$ now becomes  
much stronger (\ref{eq:6_12}). 
Efficient filtration of atmospheric wave-front distortions
is achieved primarily due to application
of reference fields of enhanced virtual symmetry. The method
takes advantage of using very large telescopes. Thus, the total error of
observations (atmospheric and photon noise) for a 10~m telescope 
and non-corrected images is expected to be about 10--20~$\mu$as per
10~min exposure, providing other sources of errors are small.
In some cases, at least, for the study of extra-solar planets,
such an accuracy is acceptable. 
Of course, very high precision ground-based observations can be
performed only in narrow fields; global data are to be obtained from space.
We hope the method described is
applicable in practice and that the results of this study will be
of interest.